\documentstyle[onecolumn,psfig,amsmath,times]{mn}

\title[General Relativistic Electromagnetic Fields of a
	Slowly Rotating Magnetized Neutron Star]{General
	Relativistic Electromagnetic Fields of a Slowly
	Rotating Magnetized Neutron Star. I. Formulation
	of the equations.}

\author[L.~Rezzolla, B.~J.~Ahmedov, and J.~C.~Miller]
	{L.~Rezzolla$^{(1)}$, B.~J.~Ahmedov$^{(2),\;(3)}$, and
	J.~C.~Miller$^{(1),\;(4)}$ 				\\ 
								\\
	$^1$SISSA, International School for Advanced Studies,
	Via Beirut 2-4, 34013 Trieste, Italy			\\ 
	$^2$AS-ICTP, The Abdus Salam International Centre
	for Theoretical Physics, 
	34014 Trieste, Italy 					\\
	$^3$Institute of Nuclear Physics,			
	Ulughbek, Tashkent 702132, Uzbekistan			\\
	$^4$Nuclear and Astrophysics Laboratory,
	University of Oxford, Keble Road, Oxford OX1 3RH	\\}

\pagerange{\pageref{firstpage}--\pageref{lastpage}}
\pubyear{2000}

\begin{document}

\maketitle

\label{firstpage}

\begin{abstract}
\noindent We present analytic solutions of Maxwell
equations in the internal and external background
spacetime of a slowly rotating magnetized neutron star.
The star is considered isolated and in vacuum, with a
dipolar magnetic field not aligned with the axis of
rotation. With respect to a flat spacetime solution,
general relativity introduces corrections related both to
the monopolar and the dipolar parts of the gravitational
field. In particular, we show that in the case of
infinite electrical conductivity general relativistic
corrections due to the dragging of reference frames are
present, but only in the expression for the electric
field.  In the case of finite electrical conductivity,
however, corrections due both to the spacetime curvature
and to the dragging of reference frames are shown to be
present in the induction equation. These corrections
could be relevant for the evolution of the magnetic
fields of pulsars and magnetars.  The solutions found,
while obtained through some simplifying assumption,
reflect a rather general physical configuration and could
therefore be used in a variety of astrophysical
situations.
\end{abstract}

\begin{keywords}
relativity -- (magnetohydrodynamics) MHD -- stars:
neutron -- rotation -- magnetic fields
\end{keywords} 

\date{Accepted 0000 00 00.
      Received 0000 00 00.}

\section{Introduction}
	The investigation of the influence of strongly
curved spacetimes on the properties of electromagnetic
fields has an interest of its own which is further
increased when these effects could be related to a rich
observable phenomenology. This coupling between general
relativistic effects and electromagnetic fields is
expected to be particularly important in the vicinity of
neutron stars which are among the most relativistic
astrophysical objects and are characterized by very
intense magnetic fields (Lamb 1991, Glendenning 1996). A
number of different observations indicate that in young
neutron stars the surface magnetic field strengths are of
the order of \hbox{$10^{11}-10^{13}$ G.}  In some
exceptional cases, as those of magnetars, magnetic field
strengths \hbox{$\ge 5 \times 10^{14}$ G} are considered
responsible for the phenomenology observed in soft
gamma-ray repeaters (Duncan \& Thompson 1992, Thompson \&
Duncan 1995). Older neutron stars, observed as recycled
pulsars and low mass X-ray binaries, show instead surface
magnetic fields that are much weaker $\le 10^{10}$ G
suggesting that these are subject to a decay, even if it
is still difficult to establish whether the decay is due
to accretion (Geppert \& Urpin, 1994; Konar \&
Bhattacharya, 1997) or to other processes.

	In the case of isolated neutron stars, the
possibility of magnetic field decay as a result of
accretion does not arise, but there are still a number of
different ways in which the energy stored in the magnetic
can be lost. This can happen either through the emission
of electromagnetic (dipole) radiation, through Ohmic
decay, through ambipolar diffusion, or through more
complicated effects such as ``Hall cascades'' (see
Goldreich and Reisenneger 1992 for a review). The
investigation of these scenarios requires combined
efforts. On one hand, there is the search for a more
precise description of the microphysics of the processes
involved, some of which are still not well quantified. On
the other hand, attention is paid to a more realistic
description of the gravitational effects on the
properties of the electromagnetic fields in highly curved
spacetimes and this is also the motivation of this work.

	The investigation of the general relativistic
corrections to the solution of Maxwell equations in the
spacetime of a relativistic star has a long history. The
initial works of Ginzburg \& Ozernoy (1964), Anderson \&
Cohen (1970) and of Petterson (1974) on the stationary
electromagnetic fields in a Schwarzschild spacetime have
revealed that the spacetime curvature produces magnetic
fields which are generally stronger than their Newtonian
counterparts (see also Wasserman \& Shapiro 1983 for a
subsequent derivation). Sengupta (1995) has reconsidered
this problem and also looked for a general relativistic
expression for the electric field in the Schwarzschild
background of a neutron star. As we will discuss in
Section \ref{srst_es} the method used in his derivation
is not entirely correct and the results obtained for the
electric field are not solutions of Maxwell
equations. More recently, Sengupta has also considered
the problem of the Ohmic decay rate in a Schwarzschild
spacetime (Sengupta, 1997). His approach is strictly
valid only for the region of spacetime external to the
star as it does not provide a correct general
relativistic description of the electromagnetic fields
internal to the star. Within these approximations,
however, Sengupta (1997) has pointed out that the effects
of intense gravitational field seem to decrease the
overall decay rate by a couple of orders of
magnitude. The same problem has also been considered in
more detail by Geppert, Page and Zannias (2000). Their
analysis was aimed at a mathematically consistent
solution of Maxwell equations also in the spacetime
region internal to the star and makes therefore use of a
generic metric for a non-rotating relativistic
star. Their results, while confirming a decrease in the
typical decay time for the magnetic field, also show that
the decay time is smaller but comparable with the one
found in flat spacetime.

	The general relativistic effects induced by the
rotation of the star were first investigated by Muslimov
\& Tsygan (1992) in the slow rotation approximation. A
similar approach was also used by Muslimov and Harding
(1997) for the electromagnetic fields external to a
rotating magnetized star. Their analysis refers to a
charge filled magnetosphere and represents the
relativistic extension of the Goldreich-Julian model.
Using a different derivation, Prasanna and Gupta (1997)
have also investigated the properties of the
electromagnetic fields in the magnetosphere of a
relativistic rotating neutron star, with special
attention being paid to the dynamics of charged test
particles.

	We here extend and unify all of the above
investigations by considering the solution of Maxwell
equations in the internal and external background
spacetime of a slowly rotating magnetized relativistic
star. The star is considered isolated and in vacuum, with
a dipolar magnetic field which is not assumed aligned
with the axis of rotation. The purpose of this paper is
threefold.  Firstly, we want to extend previous results
to the most general case of a misaligned rotator,
providing for this case also the form of the electric
field. Secondly, we want to discuss the possible role
played by frame dragging effects in the Ohmic decay for
an isolated neutron star and estimate its
importance. Thirdly, we wish to clarify a few important
aspects of the solution of Maxwell equations in the
gravitational field of a relativistic star that, when
overlooked, have led to incorrect solutions (Sengupta
1995, Prasanna and Gupta 1997). Finally, by providing a
rather general solution to the problem (although
truncated at the lowest order in the expansion of the
angular dependence) we offer a compact reference from
which all of the previous results can be easily found in
the appropriate limits and which could have practical
astrophysical applications.

	The paper is organized as follows: in Section
\ref{meq} we write the general relativistic Maxwell
equations in the metric of a slowly rotating star and the
form they assume when the electromagnetic fields are
those measured in the orthonormal frame of zero angular
momentum observers. In Section \ref{ss} we find the
stationary solutions (i.e. solutions in which the
infinite conductivity of the medium prevents a variation
in time of the star's magnetic moment) to Maxwell
equations outside and inside the misaligned rotating
star. For this we consider first the problem in Newtonian
gravity and we then extend the results to general
relativity within the slow rotation approximation.
Section \ref{nss} is devoted to the equivalent problem,
but in the case in which the magnetic field is not
supposed stationary. There, we derive the basic induction
equations for the evolution of the inner stellar magnetic
field of a misaligned rotating star. Section
\ref{conclusion} contains our conclusions and the
prospects of future developments.

	A number of appendices provide further details
about some of the calculations carried out in the main
part of the paper. In particular, Appendix A summarizes
the components of the electromagnetic tensor in a
coordinate basis and in a locally orthonormal tetrad,
while Appendix B shows the derivation of the radial
eigenfunctions for the electromagnetic fields in terms of
Legendre's equation. Appendix C shows the explicit
expressions for the surface charges and currents and,
finally, Appendix D contains an alternative and
equivalent derivation of the equations for the time
evolution of magnetic field in terms a vector
potential. Throughout, we use a space-like signature
$(-,+,+,+)$ and a system of units in which $G = 1 = c$
(However, for those expressions with an astrophysical
application we have written the speed of light
explicitely.). Greek indices are taken to run from 0 to 3
and Latin indices from 1 to 3; covariant derivatives are
denoted with a semi-colon and partial derivatives with a
comma.
  
\section{Maxwell Equations In a Slowly Rotating Spacetime}
\label{meq}

	The difficulties of an analytic solution of the
Einstein-Maxwell equations in the proximity of a rotating
relativistic star inevitably force us to the use of some
approximations. The first approximation comes from
neglecting the influence of the electromagnetic field on
the metric and by solving Maxwell equations on a given,
fixed background\footnote{This is indeed a very good
approximation since even for very highly magnetic neutron
stars the electromagnetic energy density is much smaller
than the gravitational one.}. The second approximation is
in the specific form of the background metric which we
choose to be that of a stationary, axially symmetric
system truncated at the first order in the angular
velocity $\Omega$. In a coordinate system
$(ct,r,\theta,\phi)$, the ``slow rotation metric'' for a
rotating relativistic star is (see, for example, Hartle
1967, Hartle \& Thorne 1968, Landau \& Lifshitz 1971)
\begin{equation}
\label{slow_rot}
ds^2 = -e^{2 \Phi(r)} dt^2 + e^{2
	\Lambda(r)}dr^2 - 2 \omega (r) r^2\sin^2\theta dt d\phi +
	r^2 d\theta ^2+ r^2\sin^2\theta d\phi ^2 \ ,
\end{equation}
where $\omega(r)$ can be interpreted as the angular
velocity of a free falling (inertial) frame and is also
known as the Lense-Thirring angular velocity. The radial
dependence of $\omega$ in the region of spacetime
internal to the star has to be found as the solution of
the differential equation
\begin{equation}
\label{omg_r_int}
\frac{1}{r^3} \frac{d}{dr}
	\left(r^4 {\bar j} \frac{d {\bar \omega}}{dr}\right)
	+ 4 \frac{d {\bar j}}{dr} {\bar \omega} = 0 \ ,
\end{equation} 
where we have defined 
\begin{equation}
\label{jbar}
{\bar j} \equiv e^{-(\Phi + \Lambda)} \ , 
\end{equation} 
and where 
\begin{equation}
\label{omegabar}
{\bar\omega} \equiv \Omega -\omega \ ,
\end{equation} 
is the angular velocity of the fluid as measured from the
local free falling (inertial) frame. In the vacuum region
of spacetime external to the star, on the other hand,
$\omega(r)$ is given by the simple algebraic expression
\begin{equation}
\label{omg_r_ext}
\omega (r)\equiv \frac{d\phi}{dt}=-\frac{g_{0\phi}}{g_{\phi\phi}}=
	\frac{2J}{r^3} \ ,
\end{equation} 
where $J=I(M,R)\Omega$ is the total angular momentum of
metric source as measured from infinity and $I(M,R)$ its
momentum of inertia (see Miller 1977 for a discussion of
$I$ and its numerical calculation). Outside the star, the
metric (\ref{slow_rot}) is completely known and explicit
expressions for the other metric functions are given by
\begin{equation}
e^{2 \Phi(r)} \equiv \left(1-\frac{2 M}{r}\right) 
	= e^{-2 \Lambda(r)} \ , \hskip 1.0cm
	r > R \ ,
\end{equation}
where $M$ and $R$ are the mass and radius of the star as
measured from infinity.

	An important aspect, often overlooked in the
literature, should now be underlined. The metric
(\ref{slow_rot}) is the simplest metric that provides all
of the most important general relativistic corrections to
the solution of the Maxwell equations in the
gravitational field of a rotating relativistic star. The
use of a Schwarzschild metric in place of
(\ref{slow_rot}) (Sengupta 1995, 1997) is potentially
very dangerous. Firstly, and as pointed out by Geppert et
al. (2000), a Schwarzschild metric allows for a proper
treatment of the electromagnetic fields only in the
spacetime region external to the star and leaves unsolved
the problem of a matching of the external electromagnetic
fields with the internal ones. Secondly, and despite
different claims (Sengupta 1997), a Schwarzschild metric
is intrinsically inadequate to describe physical systems
such as pulsars in which the coupling of electromagnetic
fields and rotation is a key feature. Note, on the other
hand, that using the slow-rotation approximation gives
rather accurate results for all pulsar periods so far
observed. The metric (\ref{slow_rot}) has coefficients
each of which is the lowest-order term of a series
expansion in ascending powers of $\Omega$. Comparing the
magnitude of the neglected higher order terms with that
of the one retained in each case, gives ratios of the
order $R^3\Omega^2/GM$ which is smaller than 10\% even
for the fastest-known millisecond pulsar PSR 1937+214.

	The general form of the first pair of general
relativistic Maxwell equations is given by
\begin{equation}
\label{maxwell_firstpair}
3! F_{[\alpha \beta, \gamma]} =  2 \left(F_{\alpha \beta, \gamma }
	+ F_{\gamma \alpha, \beta} + F_{\beta \gamma,\alpha}  
	\right) = 0 \ .
\end{equation}
where $F_{\alpha \beta}$ is the electromagnetic field
tensor expressing the strict connection between the
electric and magnetic four-vector fields $E^{\alpha},\
B^{\alpha}$. For an observer with four-velocity
$u^{\alpha}$, the covariant components of the
electromagnetic tensor are given by (Lichnerowicz 1967;
Ellis 1973)
\begin{equation}
\label{fab_def}
F_{\alpha\beta} \equiv 2 u_{[\alpha} E_{\beta]} +
	\eta_{\alpha\beta\gamma\delta}u^\gamma B^\delta \ .
\end{equation}
where $T_{[\alpha \beta]} \equiv \frac{1}{2}(T_{\alpha
\beta} - T_{\beta \alpha})$ and
$\eta_{\alpha\beta\gamma\delta}$ is the pseudo-tensorial
expression for the Levi-Civita symbol $\epsilon_{\alpha
\beta \gamma \delta}$ (Stephani 1990)
\begin{equation}
\eta^{\alpha\beta\gamma\delta}=-\frac{1}{\sqrt{-g}}
	\epsilon_{\alpha\beta\gamma\delta} \ ,
	\hskip 2.0cm 
\eta_{\alpha\beta\gamma\delta}=
	\sqrt{-g}\epsilon_{\alpha\beta\gamma\delta} \ , 
\end{equation}
with $g\equiv {\rm
det}|g_{\alpha\beta}|=-e^{2(\Phi+\Lambda)} r^4
\sin^2\theta$ for the metric (\ref{slow_rot}).  A useful
class of observers is represented by the ``zero angular
momentum observers'' or ZAMOs (Bardeen, Press \&
Teukolsky 1972). These are observers that are locally
stationary (i.e. at fixed values of $r$ and $\theta$) but
who are ``dragged'' into rotation with respect to a
reference frame fixed with respect to distant
observers. At first order in $\Omega$ they have
four-velocity components given by
\begin{equation}
\label{uzamos}
(u^{\alpha})_{_{\rm ZAMO}}\equiv 
	e^{-\Phi(r)}\bigg(1,0,0,\omega\bigg) \ ;
	\hskip 2.0cm
(u_{\alpha})_{_{\rm ZAMO}}\equiv 
	e^{\Phi(r)}\bigg(- 1,0,0,0 \bigg) \ .
\end{equation} 

	In the coordinate system $(ct,r,\theta,\phi)$ and
with the definition (\ref{fab_def}) referred to the
observers (\ref{uzamos}), the first pair of Maxwell
equations (\ref{maxwell_firstpair}) take then the form
(see Appendix A for the explicit expressions of the
electromagnetic tensor)
\begin{equation}
\label{max1_ea}
\left( e^{\Lambda}r^2\sin\theta B^i \right)_{,i}=0 \ ,
\end{equation}
\begin{eqnarray}
\label{max1_eb}
&& \left(e^{\Lambda}r^2\sin\theta \right)\frac{\partial B^{r}}{\partial t} 
	= {e^\Phi}\left(E_{\theta,\phi}- E_{\phi ,\theta} \right)
	-\left({\omega} e^{\Lambda}r^2\sin\theta
	\right)B^{r}_{\ ,\phi} \ ,
\\
\label{max1_ec}
&& \left(e^{\Lambda}r^2\sin\theta \right)
	\frac{\partial B^{\theta}}{\partial t}
	= \left( E_{\phi} \  {e^\Phi}\right)_{,r}-  {e^\Phi}E_{r,\phi}
	-\left({\omega} e^{\Lambda}r^2\sin\theta \right)
	B^{\theta}_{\ ,\phi}\ ,
\\
\label{max1_ed}
&& \left(e^{\Lambda}r^2\sin\theta \right)\frac{\partial B^{\phi}}{\partial t} 
	= -\left( E_{\theta} \ {e^{\Phi}}\right)_{,r}+
	{e^{\Phi}}E_{r,\theta}+\sin\theta
	\left( {\omega} e^{\Lambda}r^2 B^{r}\right)_{,r}
	+{\omega} e^{\Lambda}r^2
	\left( \sin\theta B^{\theta}\right)_{,\theta} \ . 
\end{eqnarray}	

	The general form of the second pair of Maxwell
equations is given by
\begin{equation}
\label{maxwell_secondpair}
F^{\alpha \beta}_{\ \ \ \ ;\beta} = 4\pi J^{\alpha} 
\end{equation}
where the four-current $J^{\alpha}$ is a sum of
convection and conduction currents
\begin{equation}
J^{\alpha}=\rho_e w^\alpha + j^\alpha \ ,
	\hskip 2.0cm j^\alpha w_\alpha \equiv 0 \ ,
\end{equation}
with ${\boldsymbol w}$ being the conductor four-velocity
and $\rho_e$ the proper charge density. If the conduction
current $j^\alpha$ is carried by the
electrons\footnote{This is a reasonable assumption if the
neutron star has a temperature such that the atomic
nuclei are frozen into a lattice and the electrons form a
completely relativistic, and degenerate gas.}  with
electrical conductivity $\sigma$, Ohm's law can then be
written as
\begin{equation}
\label{ohm}
j_\alpha = \sigma F_{ \alpha \beta}w^\beta \ ,
\end{equation}
while a more general expression can be found in Ahmedov
(1999). We can now rewrite the second pair of Maxwell
equations as
\begin{eqnarray}
\label{max2_ea}
\left({e^{\Lambda}r^2\sin\theta} E^i\right)_{,i}
	&=& 4\pi e^{\Phi+\Lambda}r^2\sin\theta J^0 \ ,
\\
\label{max2_eb}
e^{\Phi} \left( B_{\phi\;,\theta} -  B_{\theta\;,\phi} \right)
	-\left( {\omega}e^{\Lambda}r^2\sin\theta\right) 
	E^r_{\ \;,\phi}
	&=& \left(e^{\Lambda}r^2\sin\theta \right)
	\frac{\partial E^r}{\partial t}
	+4\pi e^{\Phi+\Lambda}r^2\sin\theta J^r\ , 
\\
\label{max2_ec}
{e^{\Phi}}B_{r,\phi}-\left({e^\Phi} B_\phi \right)_{,r}
	-\left( {\omega}e^{\Lambda}r^2\sin\theta\right)
	E^\theta_{\;,\phi}
	&=& \left(e^{\Lambda}r^2\sin\theta
	\right)\frac{\partial E^\theta}{\partial t}
	+4\pi e^{\Phi+\Lambda}r^2\sin\theta J^\theta \ ,
\\
\label{max2_ed}
\left({e^{\Phi}} B_\theta \right)_{,r} - e^{\Phi}B_{r\;,\theta}
	+\sin\theta\left({\omega} e^{\Lambda}r^2 E^r\right)_{,r}
	+{\omega} e^{\Lambda}r^2\left(\sin\theta 
	E^\theta\right)_{,\theta}
	&=& \left(e^{\Lambda}r^2\sin\theta \right)
	\frac{\partial E^\phi}{\partial t} + 
	4\pi e^{\Phi+\Lambda}r^2\sin\theta J^\phi \ .
	\hskip 1.5 cm
\end{eqnarray}

	Maxwell equations assume a familiar
flat-spacetime form when projected onto a locally
orthonormal tetrad. In principle such tetrad is
arbitrary, but in the case of a relativistic rotating
metric source a ``natural'' choice is offered by the
tetrad carried by the ZAMOs. Using (\ref{uzamos}) we find
that the components of the tetrad $\{{\boldsymbol
e}_{\hat \mu}\} = ({\boldsymbol e}_{\hat 0}, {\boldsymbol
e}_{\hat r}, {\boldsymbol e}_{\hat \theta}, {\boldsymbol
e}_{\hat \phi})$ carried by a ZAMO observer are
\begin{eqnarray}
\label{zamo_tetrad_0}
&&{\boldsymbol e}_{\hat 0}^{\alpha}  =  
	e^{-\Phi}\bigg(1,0,0,{\omega}\bigg) \ ,  	\\ 
\label{zamo_tetrad_1}
&&{\boldsymbol e}_{\hat r}^{\alpha}  =  
	e^{-\Lambda}\bigg(0,1,0,0\bigg) \ ,		\\ 
\label{zamo_tetrad_2}
&&{\boldsymbol e}_{\hat \theta}^{\alpha}  =  
	\frac{1}{r}\bigg(0,0,1,0\bigg)	\ , 		\\ 
\label{zamo_tetrad_3}
&&{\boldsymbol e}_{\hat \phi}^{\alpha}  =  
	\frac{1}{r\sin\theta}\bigg(0,0,0,1\bigg) \ . 		 
\end{eqnarray}

\noindent The 1-forms $\{{\boldsymbol \omega}^{\hat
\mu}\} = ({\boldsymbol \omega}^{\hat 0}, {\boldsymbol
\omega}^{\hat r}, {\boldsymbol \omega}^{\hat \theta},
{\boldsymbol \omega}^{\hat \phi})$, corresponding to this
tetrad have instead components
\begin{eqnarray}
\label{zamo_tetrad_1-forms_0}
&&\boldsymbol{\omega}^{\hat 0}_{\alpha}  =  
	e^{\Phi}\bigg(1,0,0,0\bigg)\ ,			\\ 
\label{zamo_tetrad_1-forms_1}
&&\boldsymbol{\omega}^{\hat r}_{\alpha}  =  
	e^{\Lambda}\bigg(0,1,0,0\bigg)\ , 		\\ 
\label{zamo_tetrad_1-forms_2}
&&\boldsymbol{\omega}^{\hat \theta}_{\alpha}  =  
	{r}\bigg(0,0,1,0\bigg)\ , 			\\ 
\label{zamo_tetrad_1-forms_3}
&&\boldsymbol{\omega}^{\hat \phi}_{\alpha}  =  
	{r\sin\theta}\bigg(-{\omega},0,0,1\bigg)\ .
\end{eqnarray}

	We can now rewrite Maxwell equations
(\ref{max1_ea})--(\ref{max1_ed}) and
(\ref{max2_ea})--(\ref{max2_ed}) in the ZAMO reference
frame by contracting (\ref{maxwell_firstpair}) and
(\ref{maxwell_secondpair}) with
(\ref{zamo_tetrad_0})--(\ref{zamo_tetrad_3}) and
(\ref{zamo_tetrad_1-forms_0})--(\ref{zamo_tetrad_1-forms_3}).
After some lengthy but straightforward algebra, we obtain
Maxwell equations in the more useful form
\begin{equation}
\label{max1a}
\sin\theta \left(r^2B^{\hat r}\right)_{,r}+
	e^{\Lambda}r\left(\sin\theta B^{\hat \theta}\right)_{,\theta} + 
	e^{\Lambda} r B^{\hat \phi}_{\ , \phi} = 0 \ ,
\end{equation}
\begin{eqnarray}
\label{max1b}
\left({r\sin\theta}\right)\frac{\partial B^{\hat r}}{\partial t}
	& = & {e^\Phi} \left[E^{\hat\theta}_{\ ,\phi}- \left(\sin\theta 
	E^{\hat \phi} \right)_{,\theta}\right]
	- \left({{\omega} r\sin\theta}\right)B^{\hat r}_{\ ,\phi} \ ,
\\
\label{max1c}
\left({e^{\Lambda}r\sin\theta}\right)
	\frac{\partial B^{\hat \theta}}{\partial t}
	&=& -e^{\Phi+\Lambda} E^{\hat r}_{\ ,\phi} + 
	\sin\theta \left(r e^{\Phi} E^{\hat \phi} \right)_{,r}
	- \left({{\omega}e^{\Lambda}r\sin\theta}\right) 
	B^{\hat \theta}_{\ ,\phi} \ ,
\\
\label{max1d}
\left({e^{\Lambda}r}\right)
	\frac{\partial B^{\hat \phi}}{\partial t}
	&=& - \left(r e^{\Phi} E^{\hat \theta}\right)_{,r} 
	+ e^{\Phi+\Lambda}E^{\hat r}_{ \ ,\theta}
	+ {\sin\theta}\left({\omega} r^2 B^{\hat r}\right)_{,r}
	+ {\omega} e^{\Lambda}r\left(\sin\theta 
	B^{\hat \theta}\right)_{,\theta} \ 
\end{eqnarray}

\noindent and
\begin{eqnarray}
\label{max2a}
\sin\theta\left(r^2 E^{\hat r} \right)_{,r}+
	{e^{\Lambda}r}\left(\sin\theta E^{\hat \theta}\right)_{,\theta}
	+ e^{\Lambda}r E^{\hat \phi}_{\;,\phi}
	& = & {4\pi e^{\Lambda}}r^2\sin\theta J^{\hat t}\ , 
\\
\label{max2b}
e^{\Phi}\left[\left(\sin\theta  B^{\hat \phi} \right)_{,\theta}
	- B^{\hat\theta}_{\ ,\phi}\right]- 
	\left({{\omega} r\sin\theta }\right)E^{\hat r}_{\;,\phi}
	& = &\left({r\sin\theta}\right)
	\frac{\partial E^{\hat r}}{\partial t}
	+{4\pi}e^{\Phi}r\sin\theta J^{\hat r} \ ,
\\
\label{max2c}
e^{\Phi+\Lambda}B^{\hat r}_{\ ,\phi} - \sin\theta \left(r \ e^{\Phi} 
	B^{\hat \phi} \right)_{,r}
	-\left({{\omega} e^{\Lambda}r\sin\theta}\right)
	E^{\hat \theta}_{\;,\phi}
	& = & \left({e^{\Lambda}r\sin\theta}\right)
	\frac{\partial E^{\hat\theta}}{\partial t}
	+{4\pi e^{\Phi+\Lambda}}r\sin\theta J^{\hat\theta} \ ,
\\
\label{max2d}
\left(e^{\Phi}r B^{\hat \theta} \right)_{,r} - e^{\Phi+\Lambda}
	B^{\hat r}_{\ ,\theta} + {\sin\theta}\left({\omega} 
	r^2E^{\hat r}\right)_{,r} +
	{{\omega} e^{\Lambda}r}
	\left(\sin\theta E^{\hat \theta}\right)_{,\theta}
	& = &\left({e^{\Lambda}r}\right)
	\frac{\partial E^{\hat\phi}}{\partial t} 
	+{4\pi e^{\Phi+\Lambda}}rJ^{\hat\phi} 
	+{4\pi e^{\Lambda}}\omega r^2\sin\theta J^{\hat t} \ .
\end{eqnarray}

	Equations (\ref{max2b})--(\ref{max2d}) can now be
rewritten in a more convenient form. Taking our conductor
to be the star with four-velocity components
\begin{equation}
\label{vel}
w^\alpha \equiv e^{-\Phi(r)}
	\bigg(1,0,0,{\Omega}\bigg) \ ,
\hskip 2.0cm
w_\alpha \equiv 
	e^{\Phi(r)} \bigg(-1,0,0,
	\frac{\bar{\omega} r^2\sin ^2\theta}{e^{2\Phi(r)} }\bigg)\ ,
\end{equation}
we can use Ohm's law (\ref{ohm}) to derive the following
explicit components of $J^{\hat \alpha}$ in the ZAMO
frame
\begin{eqnarray}
\label{current1}
&&J^{\hat t} = {\rho_e}+
	\sigma\frac{\bar{\omega}r\sin\theta}
	{e^{\Phi}}E^{\hat\phi}\ , 
\\\nonumber\\
\label{current2}
&&J^{\hat r} = \sigma
	\left( E^{\hat r} -
	\frac{\bar{\omega} r \sin\theta }{e^{\Phi}}
	B^{\hat \theta}\right)\ , 
\\\nonumber\\
\label{current3}
&&J^{\hat\theta} = \sigma\left(E^{\hat \theta}+
	\frac{\bar{\omega} r \sin\theta }{e^{\Phi}}
	B^{\hat r}\right)\ ,
\\\nonumber\\
\label{current4}
&&J^{\hat\phi} = {\sigma E^{\hat \phi}} + 
	\frac{\bar{\omega}r\sin\theta}{e^{\Phi}}\rho_e\ .
\end{eqnarray}

\noindent Next, we discuss a few assumptions that are
going to be used hereafter. Firstly, we assume there is
no matter outside the star so that the conductivity
$\sigma=0$ for $r > R$ and that $\sigma \neq 0$ only in a
shell with $R_{_{IN}}\le r \le R$ (e.g. the neutron star
crust). Secondly, we consider $\sigma$ to be uniform
within this shell (Note that this might be incorrect in
the outermost layers of the neutron star but is a rather
good approximation on the crust as a whole.). Thirdly, we
ignore the contributions coming from displacement
currents. The latter could, in principle, be relevant in
the evolution of the electromagnetic fields, but their
effects are negligible on timescales that are long as
compared with the electromagnetic waves crossing time. In
view of this, we will neglect in
(\ref{max2b})--(\ref{max2d}) all terms involving time
derivatives of the electric field and use Ohm's law to
rewrite equations (\ref{max2b}) and (\ref{max2c}) as
\begin{eqnarray}
\label{rel1}
&& {r\sin\theta E^{\hat r}} = 
	\frac{1}{4\pi\sigma}\left[
	\left(\sin\theta B^{\hat \phi}\right)_{\;,\theta}
	- B^{\hat\theta}_{\; ,\phi}\right] + 
	{\mathcal O} (\Omega) \ ,  
\\
\label{rel2}
&& {e^{\Lambda}r\sin\theta E^{\hat\theta}} =
	\frac{e^{-\Phi}}{4\pi\sigma }\left[
	e^{\Phi+\Lambda}B^{\hat r}_{\ ,\phi}- \sin\theta 
	\left(r e^{\Phi} B^{\hat \phi} \right)_{,r}\right] 
	+ {\mathcal O} (\Omega) \ .
\end{eqnarray}
Substituting (\ref{rel1}) and (\ref{rel2}) in the left
hand sides of equations (\ref{max2b})--(\ref{max2d})
eliminates the dependence from the electric field and
yields
\begin{eqnarray}
\label{max3b}
\left[\left(\sin\theta B^{\hat \phi} \right)_{,\theta}
	- B^{\hat\theta}_{\ ,\phi}\right]- \frac{{\omega} 
	e^{-\Phi}}{4\pi\sigma}\left[
	\left(\sin\theta  B^{\hat \phi} \right)_{,\theta}
	- B^{\hat\theta}_{\ ,\phi}\right]_{,\phi}
	&=& {4\pi}r\sin\theta J^{\hat r} \ ,  
\\
\label{max3c}
e^{\Phi+\Lambda}B^{\hat r}_{\ ,\phi} - \sin\theta \left(e^{\Phi}r 
	B^{\hat \phi} \right)_{,r}
	-\frac{{\omega}e^{-\Phi}}{4\pi\sigma}
	\left[e^{\Phi+\Lambda}B^{\hat r}_{\ ,\phi} - 
	\sin\theta \left(e^{\Phi}r B^{\hat \phi} \right)_{,r}
	\right]_{,\phi} 
	&=& {4\pi e^{\Phi+\Lambda}}r\sin\theta J^{\hat\theta} \ ,  
\\
\label{max3d}
\left(e^{\Phi}r B^{\hat \theta} \right)_{,r} - e^{\Phi+\Lambda}
	B^{\hat r}_{\ ,\theta} + \frac{1}{4\pi}\left\{\frac{{\omega}r}
	{\sigma}\left[\left(\sin\theta  B^{\hat \phi} \right)_{,\theta}
	- B^{\hat\theta}_{\ ,\phi}\right]\right\}_{,r}+
	\hskip 2.0cm & &
\nonumber\\ 
	\frac{{\omega}e^{-\Phi}}{4\pi\sigma}
	\left[e^{\Phi+\Lambda}B^{\hat r}_{\ ,\phi} - 
	\sin\theta \left(e^{\Phi}r B^{\hat \phi} \right)_{,r}
	\right]_{,\theta}
	&=& {4\pi e^{\Phi+\Lambda}} rJ^{\hat\phi} 
	+{4\pi e^{\Lambda}}\omega r^2\sin\theta \rho_e \ .  
\end{eqnarray}

\section{Stationary Solutions to Maxwell Equations}
\label{ss}

	In this Section we will look for stationary
solutions of the Maxwell equation, i.e. for solutions in
which we assume that the magnetic moment of the magnetic
star does not vary in time as a result of the infinite
conductivity of the medium. Note that this does not mean
the electromagnetic fields are independent of time. As a
result of the misalignment between the magnetic dipole
${\boldsymbol \mu}$ and the angular velocity vector
${\boldsymbol \Omega}$, in fact, both the magnetic and
the electric fields will posses a {\it periodic} time
dependence produced by the precession of ${\boldsymbol
\mu}$ around ${\boldsymbol \Omega}$ (see Fig.~1).

\subsection{Rotating Magnetized Conductor in a Minkowski Spacetime}
\label{mst}
	
	Before looking at the problem of a magnetized
rotating conductor in a rotating spacetime, it is useful
to start with a simpler analogous configuration: that of
a rotating magnetized conductor in a Minkowski (flat)
spacetime. This will provide important insight for the
search of general relativistic solutions and useful
limits against which match the fully relativistic
solutions.

	Consider therefore a conducting magnetized sphere
of radius $R$ rotating at angular velocity $\Omega$, and
with the magnetic four-vector field ${\boldsymbol B}$
being uniform (in radius) inside the sphere and dipolar
outside (This is a simple but instructive
example.). Because of discontinuities in the fields
across the surface of the sphere we will refer to as {\sl
interior solutions} those solutions valid within the
radial range $R_{_{IN}} \le r \le R$, and to as {\sl
exterior solutions} those valid in the range $R < r \le
\infty$.

\subsubsection{Interior Solution}
\label{mst_is}

	The interior solution for the electromagnetic
fields of a magnetized sphere with magnetic moment
aligned with the rotation axis was found by Ruffini and
Treves in 1973 (Ruffini \& Treves 1973). Extending it to
the case of a misaligned rotator we obtain
\begin{eqnarray}
&& B^{\hat r} = \frac{2\mu} {R^3} \left(
	\cos\chi \cos\theta +
	\sin\chi \sin\theta \cos\lambda	\right)		\ ,
\\ \nonumber \\
&& B^{\hat \theta} = -\frac{2\mu}{R^3} \left(
	\cos\chi \sin\theta
	- \sin\chi \cos\theta \cos\lambda \right)	\ ,
\\\nonumber\\
&& B^{\hat \phi} = -\frac{2\mu}{R^3} \sin\chi \sin\lambda \ ,
\end{eqnarray}

\noindent where ${\boldsymbol \mu}$ is the magnetic dipole moment
of the star, $\chi$ is the inclination angle of the
magnetic moment relative to the rotation axis and
$\lambda(t) \equiv \phi - {\Omega} t$ is the instantaneous
azimuthal position (see Fig.~1).

\begin{figure}
\null\vspace{0.5cm}
\centerline{
\psfig{file=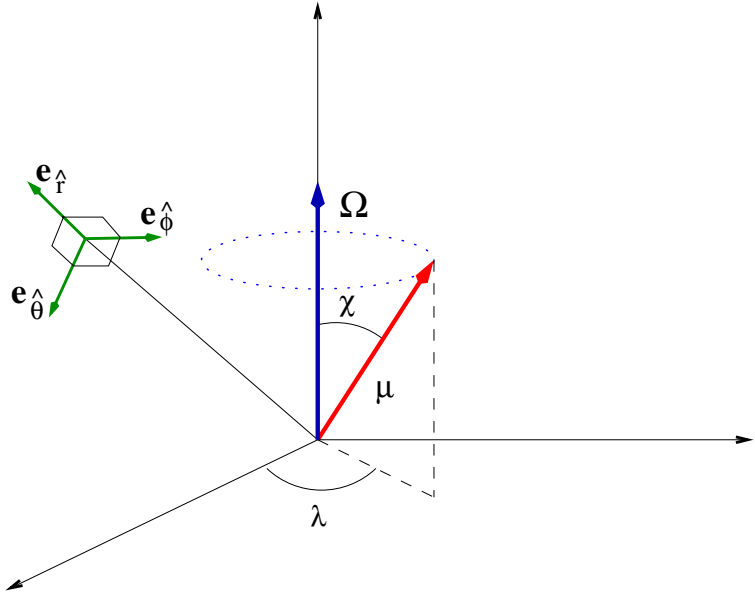,angle=0,width=10.0cm}
        }
\noindent{\small \null\vspace{0.5cm}
\addtolength{\baselineskip}{-3pt} 
\caption{\ Schematic representation of a misaligned
rotator. Here $({\boldsymbol e}_{\hat 0}, {\boldsymbol
e}_{\hat r}, {\boldsymbol e}_{\hat \theta}, {\boldsymbol
e}_{\hat \phi})$ is a local orthonormal frame,
${\boldsymbol \mu}$ is the magnetic dipole moment of the
star, $\chi$ is the inclination angle relative to the
rotation axis, and $\lambda$ the instantaneous azimuthal
position.}}
\vspace{0.5cm}
\end{figure}

	The expressions for the components of the
electric field are very simple to derive when one assumes
that the sphere is a ``perfect conductor'' (i.e. $\sigma
\rightarrow \infty$) and there are no conduction currents
inside the sphere. In this case, Ohm's law can be used to
obtain
\begin{eqnarray}
\label{ief_msta}
&& E^{\hat r} = \frac{\Omega r \sin\theta}{c}
	B^{\hat \theta} = -\frac{2\mu\Omega r \sin\theta}{cR^3}
	\left(\cos\chi \sin\theta
	- \sin\chi \cos\theta \cos\lambda \right)
	\ , 
\\
\label{ief_mstb}
&& E^{\hat \theta} = - \frac{\Omega r \sin\theta}{c}
	B^{\hat r} = - \frac{2\mu\Omega r \sin\theta}{cR^3}
	\left(\cos\chi \cos\theta +
	\sin\chi \sin\theta \cos\lambda	\right)
	\ ,
\\
\label{ief_mstc}
&& E^{\hat \phi} = 0 \ .
\end{eqnarray} 

\subsubsection{Exterior Solution}
\label{mst_es}

	The solution to this problem, i.e. to the form of
the electromagnetic fields external to a misaligned
rotating magnetized sphere, was found in 1955 by Deutsch
(Deutsch, 1955). The full solutions are complicated
expressions involving spherical Bessel functions of the
third kind, but they become much simpler when truncated
at the lowest order. The magnetic field components, in
particular, have the form
\begin{eqnarray}
\label{bd_1}
&& B^{\hat r} = \frac{2 \mu} {r^3} \left(
	\cos\chi \cos\theta +
	\sin\chi \sin\theta \cos\lambda	\right)		\ ,
\\ \nonumber \\
\label{bd_2}
&& B^{\hat \theta} = \frac{\mu}{r^3} \left(
	\cos\chi \sin\theta
	- \sin\chi \cos\theta \cos\lambda \right)	\ ,
\\\nonumber\\
\label{bd_3}
&& B^{\hat \phi} = \frac{\mu}{r^3} \sin\chi \sin\lambda 	\ ,
\end{eqnarray}

\noindent while the corresponding electric fields are given by
\begin{eqnarray}
\label{ed_1}
&& E^{\hat r} = - \frac{\mu\Omega R^2}{cr^4}
	\left[\cos\chi (3 \cos^2\theta-1)+3\sin\chi\cos\lambda
	\sin\theta\cos\theta\right]\ , 
\\\nonumber \\
\label{ed_2}
&& E^{\hat \theta} = -\frac{\mu\Omega}{cr^2}\left\{\frac{2 R^2}{r^2}
	\cos\chi\sin\theta\cos\theta +\sin\chi\left[1-
	\frac{R^2}{r^2}(\cos^2\theta-\sin^2\theta)\right]
	\cos\lambda\right\} \ ,
\\\nonumber \\
\label{ed_3}
&& E^{\hat \phi} = \frac{\mu\Omega}{cr^2}\sin\chi\cos\theta\sin\lambda
	\left(1-\frac{R^2}{r^2}\right) \ .
\end{eqnarray}

\noindent Three interesting features of solutions
(\ref{bd_1})--(\ref{bd_3}) and (\ref{ed_1})--(\ref{ed_3})
should be noticed. The first one is given by the periodic
time modulation introduced by the precession of the
magnetic moment and which disappears when the dipole is
aligned, i.e. for $\chi=0$. The second feature is that,
as one might have expected on the basis of symmetry
considerations, the toroidal components of the external
electromagnetic fields are just a by-product of the
misalignment between the rotation axis and the magnetic
dipole and again disappear when $\chi=0$. Finally, the
third relevant feature is the appearance of an electric
field of ${\mathcal O}(\Omega)$ introduced by the
rotation of the sphere and whose quadrupolar part
[i.e. $\propto (3 \cos^2\theta-1)$] is present also in
the case of an aligned rotator. As we will see in Section
\ref{srst_es}, where we study the analogous problem in a
slowly rotating spacetime, an additional contribution of
${\mathcal O}(\omega)$\footnote{Hereafter we will refer
to as ${\mathcal O}(\omega)$ any quantity that is the
result of the dragging of reference frames and that is
therefore $\propto g_{0\phi}$.} to the form of the
external electric field will be introduced by the general
relativistic frame dragging effect\footnote{Note that an
electric field induced by the rotation of the star must
appear also in the general relativistic case. This is not
present in the solution proposed by Prasanna and Gupta
1997, where the external electric field is only of
${\mathcal O}(\omega)$ and the radial dependence does not
contain higher order terms in $M/r$.}.

\subsection{Rotating Magnetized Conductor in a Slowly
	Rotating Spacetime}
\label{srst}

	We now consider the general relativistic analogue
of the problem in Section \ref{mst} and look for a
solution of Maxwell equations
(\ref{max1a})--(\ref{max1d}) and
(\ref{max2a})--(\ref{max2d}) assuming that magnetic field
of the star is dipolar. To simply the search for a
solution we look for separable solutions of Maxwell
equations in the form
\begin{eqnarray}
\label{ansatz_1}
&& B^{\hat r}(r,\theta,\phi,\chi,t) = F(r)\Psi_1(\theta,\phi,\chi,t)\ ,
\\\nonumber\\ 
\label{ansatz_2}
&& B^{\hat \theta}(r,\theta,\phi,\chi,t) = G(r)\Psi_2(\theta,\phi,\chi,t)\ ,
\\\nonumber\\
\label{ansatz_3}
&& B^{\hat \phi}(r,\theta,\phi,\chi,t) = H(r)\Psi_3(\theta,\phi,\chi,t)\ ,
\end{eqnarray}

\noindent where $F(r), G(r)$, and $H(r)$ will account for
the relativistic corrections due to a curved background
spacetime. 

	A considerable simplification comes from the fact
that, at first order in $\Omega$, the solutions for the
electromagnetic fields will not acquire general
relativistic corrections to their angular dependence. We
therefore expect that, as for the case of the Deutsch
solution, the general expressions for the angular
eigenfunctions $\Psi_i$, with $i=1,\dots,3$, will have a
complicated angular dependence expressed in terms of
spherical Bessel functions of the third kind. This is
however over-complicated and for most of the
astrophysical applications it would sufficient to know
the form at the lowest order which can be known by
requiring that the solutions match the lowest order
solution for a misaligned rotating dipole in flat
spacetime. In this case, then, we obtain
\begin{eqnarray}
\label{psi1}
&& \Psi_1(\theta,\phi,\chi,t) = \cos\chi \cos\theta +
		\sin\chi \sin\theta \cos\lambda(t) \ ,
\\\nonumber\\ 
\label{psi2}
&& \Psi_2(\theta,\phi,\chi,t) = \cos\chi \sin\theta
		- \sin\chi \cos\theta \cos\lambda(t) \ ,
\\\nonumber\\
\label{psi3}
&& \Psi_3(\phi,\chi,t) = \sin\chi \sin\lambda(t) \ ,
\end{eqnarray}
which also satisfy the following useful relations
\begin{equation}
\Psi_{1, \theta} = - \Psi_2 \ , 		\hskip 2.0cm
\Psi_{1, \phi}   = - \Psi_3 \sin\theta \ , 	\hskip 2.0cm
\Psi_{2, \theta} =   \Psi_1 \ , 		\hskip 2.0cm
\Psi_{2, \phi}   =   \Psi_3 \cos\theta\ .
\end{equation}

	Maxwell equations (\ref{max1a}),
(\ref{max3b})--(\ref{max3d}) with the ansatz
(\ref{ansatz_1})--(\ref{ansatz_3}), yield the following
set of equations
\begin{eqnarray}
\label{ans_sol_1}
\left[\left(r^2 F\right)_{, r} + 2e^{\Lambda}r G\right] 
	\sin\theta\left(\cos\chi\cos\theta +
	\sin\chi\sin\theta\cos\lambda\right)+
	e^{\Lambda}r\left(H-G\right)\sin\chi\cos\lambda 
	&=& 0 \ ,
\\\nonumber\\ 
\label{ans_sol_2}
\left(H-G\right)\cos\theta\sin\chi\left[\sin\lambda -
	\frac{\omega e^{-\Phi}}{4\pi\sigma}\cos\lambda
	\right] &=& {4\pi} r\sin\theta J^{\hat r} \ ,
\\\nonumber\\ 
\label{ans_sol_3}
\left[\left(r e^{\Phi} H\right)_{, r} 
	+ e^{\Phi+\Lambda} F\right] 
	\sin\theta\sin\chi\left[\sin\lambda -
	\frac{\omega e^{-\Phi}}{4\pi\sigma}\cos\lambda
	\right] &=& - {4\pi} e^{\Phi+\Lambda}
	r\sin\theta J^{\hat \theta} \ ,
\\ \nonumber\\
\label{ans_sol_4}
\left\{\left[
	\frac{\omega r}{4\pi\sigma}\left(H-G\right)
	\right]_{, r}-\frac{\omega r}{4\pi\sigma}
	\Phi_{, r}\left(G-H\right) 
	-\frac{\omega e^{-\Phi}}{4\pi\sigma}
	\left[\left(r e^{\Phi} H\right)_{, r} 
	+ e^{\Phi+\Lambda} F\right] \right\}
	\cos\theta\sin\chi\sin\lambda
\nonumber\\
	+ \left[\left(r e^{\Phi} G\right)_{, r} 
	+ e^{\Phi+\Lambda} F\right]\left(
	\cos\chi\sin\theta -\sin\chi\cos\theta\cos\lambda\right) 
	&=& {4\pi} e^{\Phi+\Lambda} r J^{\hat \phi} \ .
\end{eqnarray}

	Next, we will distinguish between an external
vacuum solution to Maxwell equations (for which fully
analytic solution can be given) from the interior
non-vacuum solution. Since we are treating the interior
of the star as a perfect conductor and the exterior of
the star as vacuum, we can impose $J^{\hat r} = J^{\hat
\theta} = J^{\hat \phi} = 0$ in
(\ref{ans_sol_1})--(\ref{ans_sol_4}) and obtain as
Maxwell equations for the radial part of the magnetic
field
\begin{eqnarray}
\label{ir_1}
\left(r^2 F\right)_{, r} + 2 e^{\Lambda}r G = 0 \ , && 
\\\nonumber\\ 
\label{ir_2}
\left(r e^{\Phi} H\right)_{, r}	+ e^{\Phi+\Lambda} F = 0\ , &&
\\ \nonumber\\
\label{ir_3}
H - G = 0  \ . && 
\end{eqnarray}

\noindent Note a first important result in the system of
equations (\ref{ir_1})--(\ref{ir_3}). In the case of
stationary electromagnetic fields, the general
relativistic frame dragging effect does not introduce a
correction to the radial eigenfunctions of the magnetic
fields. In other words, in the case of infinite
conductivity and as far as the magnetic field is
concerned, the study of Maxwell equations in a slow
rotation metric provides no additional information with
respect to a non-rotating metric. The frame dragging
effects are therefore expected to appear at ${\mathcal
O}(\omega^2)$.

\subsubsection{Interior solution}
\label{srst_is}
	Limiting the solution to an inner radius
$R_{_{IN}}$ removes the problem of suitable boundary
conditions for $r\rightarrow 0$, and reflects the basic
ignorance of the properties of magnetic fields in the
interior regions of neutron stars.

	It is important to notice how the system of
equations (\ref{ir_1})--(\ref{ir_3}) combines information
about the structure and physics of the star (through the
metric functions $\Phi$ and $\Lambda$) with information
about the microphysics of the magnetic field (through the
radial eigenfunctions $F$ and $G$). As a result, a
relativistic solution for the interior electromagnetic
field cannot be given independently of a self-consistent
solution of Einstein equations for the structure of the
star. In practice, to calculate a generic solution to
(\ref{ir_1})--(\ref{ir_3}), it is necessary to start with
a (realistic) equation of state and obtain a full
solution of the relativistic star. Once the latter is
known, the system of equations (\ref{ir_1})--(\ref{ir_3})
can be solved for a magnetic field which is consistent
with the star's structure and corresponds to a magnetic
configuration of some astrophysical interest.  (This is
what done, for instance, by Gupta et al. 1998 in the case
of an internal dipolar magnetic field). 

	Alternatively, one might specify a magnetic field
configuration and look for a compatible equation of state
for the stellar structure (This is a less satisfactory
way to proceed but one which is useful to get insight
into the problem.). In this case, the simplest possible
solution to the system (\ref{ir_1})--(\ref{ir_3}) is one
in which the magnetic field is constant throughout the
region of the star of interest and is therefore the
general relativistic analogue of the solution presented
in \ref{mst_is}. In this case, then
\begin{equation}
\label{umf}
F=\frac{C_1}{R^3} \mu \ , 
	\hskip 3.0cm 
G=-\frac{e^{-\Lambda}C_1}{R^3} \mu =-e^{-\Lambda}F\ , 
\end{equation}
where $C_1$ is an arbitrary constant whose value can be
determined after imposing the continuity across the star
surface of $B^{\hat r}$.

	We can now check whether the solution (\ref{umf})
is physically possible. Using (\ref{umf}) in the system
of equations (\ref{ir_1})--(\ref{ir_3}) requires that the
metric functions satisfy the condition
\begin{equation}
\label{condition}
\left(r e^{\Phi-\Lambda} \right)_{, r} -
	e^{\Phi+\Lambda} = 0 \ .
\end{equation}
Recalling now that Einstein equations for a spherical
star yield
\begin{equation}
e^{2 \Lambda(r)} = \left(1-\frac{2 m(r)}{r}\right)^{-1} 
	\ , \hskip 1.0 cm
	r \le R \ , 
\end{equation}
with $m(r)=4\pi\int^R_0 r^2 \rho(r) dr $, and $\rho(r)$
being the total energy density, we can rewrite
(\ref{condition}) as
\begin{equation}
\label{condition1}
\Phi_{,r}=e^{2\Lambda}\left(\frac{m+rm_{,r}}{r^2}\right) \ .
\end{equation}
On other hand, the solution of the Einstein equations for
the interior of a relativistic spherical star (i.e.  the
solution of the Tolmann-Oppenheimer-Volkoff equations;
Tolmann, 1939; Oppenheimer \& Volkoff, 1939) requires
that
\begin{equation}
\label{TOV}
\Phi_{,r}=e^{2\Lambda}\left(\frac{m+4\pi r^3 P}{r^2}\right) \ ,
\end{equation}
where $P$ is the isotropic pressure. The comparison of
(\ref{TOV}) with (\ref{condition1}) shows that the
general relativistic uniform magnetic field solution
\begin{eqnarray}
\label{imf_srst_1}
	&&B^{\hat r} = \frac{1}{R^3}
	\left(\cos\chi \cos\theta
	+ \sin\chi \sin\theta \cos\lambda \right)
	C_1 \mu \ ,
\\ \nonumber \\
\label{imf_srst_2}
	&&B^{\hat \theta} = -\frac{e^{-\Lambda}}{R^3}
	\left(\cos\chi \sin\theta
	- \sin\chi \cos\theta \cos\lambda \right)
	C_1 \mu \ ,
\\ \nonumber \\
\label{imf_srst_3}
	&&B^{\hat \phi} = -\frac{e^{-\Lambda}}{R^3}
	\left(\sin\chi \sin\lambda\right) 
	C_1 \mu \ ,
\end{eqnarray} 
is possible only for the (unrealistic) case of the
``stiff matter'' equation of state $P=\rho$.

	The corresponding form of the internal electric
field is also straightforward to derive in the case of no
conduction currents. In this case, in fact, Ohm's law
(\ref{current1}) and (\ref{current2}) yield the simple
expressions
\begin{eqnarray}
\label{ief_srst_1}
&&E^{\hat r} = \frac{\bar{\omega}r \sin\theta}
	{ce^{\Phi}} B^{\hat \theta}
	= -\frac{e^{-(\Phi+\Lambda)}r \sin\theta}{cR^3}
	\bar{\omega}\left(\cos\chi \sin\theta
	- \sin\chi \cos\theta \cos\lambda \right) 
	C_1 \mu \ ,
\\ \nonumber \\
\label{ief_srst_2}
	&&E^{\hat \theta} = -
	\frac{\bar{\omega} r \sin\theta }{ce^{\Phi}} B^{\hat r}
	= - \frac{e^{-\Phi} r \sin\theta }{cR^3} \bar{\omega} 
	\left(\cos\chi \cos\theta
	+ \sin\chi \sin\theta \cos\lambda \right) 
	C_1 \mu \ ,
\\ \nonumber \\
\label{ief_srst_3}
	&&E^{\hat \phi} = 0 \ ,
\end{eqnarray} 

\noindent where we have taken into account that $\rho_e =
{\mathcal O}(\omega)$ and that the contribution
proportional to $\bar{\omega} \rho_e$ is therefore of
higher order\footnote{Prasanna and Gupta (1997) have used
the assumption of infinite conductivity also for the
matter outside the neutron star. We note that their
expressions for the electric fields do not contain the
(important) contribution of $O(\Omega)$ and the radial
component does not seem to satisfy Ohm's law.}. Note
that, apart for red-shift correction proportional to
$e^{-\Phi}$, equations
(\ref{ief_srst_1})--(\ref{ief_srst_3}) are the same as
(\ref{ief_msta})--(\ref{ief_mstc}) with $\Omega$ being
replaced by the effective fluid velocity measured by a
free falling observer ${\bar \omega}$. The internal
charge density corresponding to the electrical field
(\ref{ief_srst_1})--(\ref{ief_srst_3}) can be calculated
after imposing that [cf. eq. (\ref{max2a})]
\begin{equation}
\rho_e = \frac{1}{4\pi}\left[
	\frac{e^{-\Lambda}}{r^2}
	\left(r^2 E^{\hat r} \right)_{,r}+
	\frac{1}{r\sin\theta}
	\left(\sin\theta E^{\hat \theta}\right)_{,\theta}
	\right]\ .
\end{equation}
Using now expressions
(\ref{ief_srst_1})--(\ref{ief_srst_3}), we easily obtain
\begin{equation}
\rho_e = \frac{1}{4\pi}\Biggl\{
	\left[3 e^{-\Phi} \bar{\omega}
	- \frac{e^{-\Lambda}}{r^2}
	\left(e^{-(\Phi+\Lambda)}\bar{\omega}r^3\right)_{,r}\right]
	\frac{\sin\theta}{cR^3}\left(\cos\chi \sin\theta
	- \sin\chi \cos\theta \cos\lambda \right)
	-\frac{2e^{-\Phi}}{cR^3} \bar{\omega}
	\cos\chi \Biggr\} C_1 \mu \ .
\end{equation}

\subsubsection{Exterior Solution}
\label{srst_es}

	The exterior solution for the magnetic field is
simplified by the knowledge of explicit analytic
expressions for the metric functions $\Phi$ and
$\Lambda$. In particular, after defining $N \equiv
e^{\Phi} =e^{-\Lambda}= (1 - 2M/r)^{1/2}$, the system
(\ref{ir_1})--(\ref{ir_3}) can be written as a single,
second-order ordinary differential equation for the
unknown function $F$
\begin{equation}
\label{leg_eq_sim}
\frac{d}{dr}\left[\left(1-\frac{2M}{r}\right)
	\frac{d}{dr}\left(r^2 F\right)\right] - 2F = 0 \ .
\end{equation}
\noindent Introducing now the new variable $x\equiv
1-r/M$, equation (\ref{leg_eq_sim}) can be written as
\begin{equation}
\label{leg_sol_1}
\frac{d}{dx}\left\{\left(\frac{1+x}{1-x}\right)\frac{d}{dx}
	\left[\left(1-x\right)^2 F\right]\right\}+ 2F = 0 \ .
\end{equation}
Equation (\ref{leg_sol_1}) is an example from a class of
equations which can be solved in terms of the Legendre
functions of the second kind $Q_{\ell}$ (see Appendix B
for details on the derivation of the solution). In the
case of equation (\ref{leg_sol_1}) we have $\ell=1$, and
(Jeffrey 1995)
\begin{equation}
Q_1=\frac{x}{2}\ln\left(\frac{1+x}{1-x}\right) - 1\ .
\end{equation}
The radial eigenfunctions $F(r),\ G(r)$, and $H(r)$, are
then given by
\begin{eqnarray}
\label{f_of_r}
&& F(r) = - \frac{3}{4M^3}
	\left[\ln N^2 + \frac{2M}{r}\left(1 +  \frac{M}{r}
	\right) \right] \mu
	\ ,
\\\nonumber\\ 
\label{g_of_r}
&& G(r) = \frac{3 N}{4 M^2 r}
	\left[\frac{r}{M}\ln N^2 +\frac{1}{N^2}+ 1 \right] \mu 
	\ ,
\\\nonumber\\ 
\label{h_of_r}
&& H(r) = G(r) \ ,
\end{eqnarray}
and satisfy the following boundary conditions:

{\it (i)} refer to a vanishing field at infinity, i.e.
\begin{equation}
\lim_{r \rightarrow \infty} F(r) = 0 \ , 
\hskip 5.0cm 
\lim_{r \rightarrow \infty} G(r) = 0 \ ;
\end{equation}

{\it (ii)} reduce to a flat spacetime solution for a
dipole, i.e.
\begin{equation}
\lim_{M/r \rightarrow 0} F(r) = \frac{2 \mu}{r^3} \ , 
\hskip 5.0cm 
\lim_{M/r \rightarrow 0} G(r) = \frac{\mu}{r^3} \ . 
\end{equation}

{\it (iii)} coincide with the corresponding radial
eigenfunctions found for a Schwarzschild spacetime
(Ginzburg \& Ozernoy, 1964; Anderson \& Cohen,
1970). This is what we expected since there are no first
order contributions due to the rotation of the spacetime.

\noindent Using expressions
(\ref{f_of_r})--(\ref{h_of_r}) we can now determine the
value of the matching constant $C_1$ by requiring that
the radial magnetic field is continuous across the star
surface, i.e. that $[B^{\hat r}(r=R)]_{_{IN}} = [B^{\hat
r}(r=R)]_{_{EXT}}$. As a result, we obtain
\begin{equation}
C_1 = - \frac{3R^3}{4M^3}
	\left[\ln\left(1-\frac{2M}{R}\right)
	+\frac{2M}{R}\left(1+\frac{M}{R}\right)\right] 
	= \frac{F(R) R^3}{\mu}\ , 
\end{equation}
whose flat spacetime limit is
\begin{equation}
\lim_{M/R \rightarrow 0} C_1 = 2\ . 
\end{equation}

	Collecting all the expressions for the radial
eigenfunctions, the stationary vacuum magnetic field
external to a misaligned magnetized relativistic star is
given by
\begin{eqnarray}
\label{sol_mfe_1}
&& B^{\hat r} = - \frac{3}{4M^3}
	\left[\ln N^2 + \frac{2M}{r}\left(1 +  \frac{M}{r}
	\right) \right] (\cos\chi \cos\theta +
	\sin\chi \sin\theta \cos\lambda)\mu
	\ , 
\\\nonumber\\ 
\label{sol_mfe_2}
&& B^{\hat \theta} = \frac{3 N}{4 M^2 r}
	\left[\frac{r}{M}\ln N^2 +\frac{1}{N^2}+ 1
	\right] (\cos\chi \sin\theta
	- \sin\chi \cos\theta \cos\lambda)\mu 
	\ ,
\\\nonumber\\
\label{sol_mfe_3}
&& B^{\hat \phi} = \frac{3 N}{4 M^2 r}
	\left[\frac{r}{M}\ln N^2 +\frac{1}{N^2}+ 1
	\right](\sin\chi \sin\lambda)\mu
	\ .
\end{eqnarray}

	The search for the form of the electric field is
much more involved than for the magnetic field. However,
hereafter we will make use of the insight gained in
Section \ref{mst_es} as a guide and start the derivation
of the solution by rewriting vacuum Maxwell equations
(\ref{max1b})--(\ref{max1d}) and (\ref{max2a}) as
\begin{eqnarray}
\label{vac_1}
\frac{3\bar{\omega}r}{4M^3N}\mu \left[\ln N^2+
	\frac{2M}{r}\left(1+\frac{M}{r}\right)\right]
	\sin\chi\sin^2\theta\sin\lambda 
	&=& \left(\sin\theta E^{\hat \phi}\right)_{,\theta} 
	- E^{\hat\theta}_{\ ,\phi} \ ,
\\
\label{vac_2}
\frac{3\bar{\omega}r}{4M^2}\mu \left[\frac{r}{M}\ln N^2 
	+ \frac{1}{N^2} + 1\right]\sin\chi\sin\theta\cos\theta\sin\lambda 
	&=&  E^{\hat r}_{\ ,\phi}  
	- \sin\theta \left(r NE^{\hat \phi} \right)_{,r} \ ,
\\
\label{vac_3}
	\frac{9\omega r}{4M^3}\mu
	\left[\ln N^2 +\frac{2M}{r}\left(1+\frac{M}{r}\right)
	\right] \left(\cos\chi\cos\theta +
	\sin\chi\sin\theta\cos\lambda\right)\sin\theta
\nonumber\\  
	+ \frac{3\bar{\omega}}{4M^2}\mu
	\left[\frac{r}{M}\ln N^2 +
	\frac{1}{N^2} + 1 \right]\sin\chi\cos\lambda 
	&=& \left(rN E^{\hat \theta}\right)_{,r} 
	- E^{\hat r}_{\ ,\theta} \ , 
\\
\label{vac3}
	 N\sin\theta\left(r^2 E^{\hat r} \right)_{,r}+
	{r}\left(\sin\theta E^{\hat \theta}\right)_{,\theta}
	+ rE^{\hat \phi}_{\ ,\phi}
	&=& 0 \ , 
\end{eqnarray}

\noindent and which already indicate that the dragging of
inertial frames with angular velocity $\omega$ introduces
electric fields in the surrounding space when magnetic
fields are present. Using as a reference the solutions
(\ref{ed_1}), (\ref{ed_2}), and (\ref{ed_3}) for a
misaligned rotating sphere in Minkowski spacetime, we
look for the simplest solutions of vacuum Maxwell
equations in the form
\begin{eqnarray}
\label{eef_srst_1}
&& E^{\hat r} = \left(f_1+f_3\right)\cos\chi (3 \cos^2\theta-1)+
	\left(g_1+g_3\right)3\sin\chi\cos\lambda
	\sin\theta\cos\theta \ , 
\\\nonumber \\
\label{eef_srst_2}
&& E^{\hat \theta} = \left(f_2+f_4\right)\cos\chi\sin\theta\cos\theta 
	+\left(g_2+g_4\right)\sin\chi\cos\lambda - 
	\left(g_5+g_6\right)\left(\cos^2\theta-\sin^2\theta\right)
	\sin\chi\cos\lambda \ ,
\\\nonumber \\
\label{eef_srst_3}
&& E^{\hat \phi} = \left[g_5+g_6 - \left(g_2+g_4\right)
	\right]\sin\chi\cos\theta\sin\lambda \ ,
\end{eqnarray}

\noindent where the unknown eigenfunctions $f_1-f_4$, and
$g_1-g_6$ can be found as solutions to vacuum Maxwell
equations and have radial dependence only. Substituting
(\ref{eef_srst_1})--(\ref{eef_srst_3}) in
(\ref{vac_1})--(\ref{vac_3}) we obtain the following set
of linear differential equations
\begin{eqnarray}
\label{f1}
N\left(r^2f_1\right)_{,r} + r f_2 = 0 \ , &&
\\\nonumber\\
\label{f2}
(r N f_2 )_{,r}+6f_1 = 0 \ , &&
\\\nonumber\\
\label{f3}
N\left(r^2 f_3\right)_{,r} + r f_4 = 0 \ , &&
\\\nonumber\\
\label{f4}
(r N f_4 )_{,r}+ 6 f_3-\frac{9\omega r}{4M^3}\mu
	\left[\ln N^2+\frac{2M}{r}
	\left(1+\frac{M}{r}\right)\right]  = 0 \ , &&
\\\nonumber\\
\label{Y1}
N\left(r^2 g_1\right)_{,r} + 2 r g_5 = 0 \ , &&
\\\nonumber\\
\left( r N g_5 \right)_{,r} + 3 g_1 = 0 \ , &&
\\\nonumber\\
\label{X1}
N\left(r^2 g_3\right)_{,r} + 2 r g_6 = 0 \ , &&
\\\nonumber\\
\left(r N g_6 \right)_{,r} + 3 g_3-
	\frac{9\omega r}{8M^3}\mu \left[\ln N^2+
	\frac{2M}{r}\left(1+\frac{M}{r}\right)\right] 
	= 0 \ . &&
\end{eqnarray}

\noindent Note that both the sets of radial
eigenfunctions $f_1-f_4$, and $g_1$, $g_3$, $g_5$, $g_6$
are linearly independent, but that relations can be
written between the two sets. In particular, the
comparison of equation (\ref{f1}) with (\ref{Y1}) and of
equation (\ref{f3}) with (\ref{X1}) indicates that
\begin{equation}
g_1 = f_1\ , 		\hskip 2.0cm
g_3 = f_3\ , 		\hskip 2.0cm
g_5 = \frac{f_2}{2}\ , 	\hskip 2.0cm
g_6 = \frac{f_4}{2}\ . 
\end{equation}

	We start the search for explicit expressions for
the radial eigenfunctions by combining equations
(\ref{f1}) and (\ref{f2}) to obtain a single differential
equation of second order for the unknown function $f_1$
\begin{eqnarray}
\label{f1_eq}
\frac{d}{dr}\left[\left(1-\frac{2M}{r}\right)\frac{d}{dr}
	\left(r^2f_1\right)\right]-6f_1=0 \ ,
\end{eqnarray}
and which can again be recast in a form similar to
equation (\ref{leg_sol_1}). Proceeding in a way analogous
to what done for the magnetic field (see Appendix B for
details) it is possible to realize that the solution
should be expressed in terms of a Legendre function of
the second kind and of order $\ell=2$. Recalling now that
(Jeffrey 1995)
\begin{equation}
Q_2(x)=\frac{1}{4}\left(3x^2-1\right)
	\ln\left(\frac{x+1}{x-1}\right) -\frac{3x}{2} \ ,
\end{equation}
we obtain, as solution to (\ref{f1_eq}) at the $\ell=2$
order in the expansion
\begin{equation}
\label{f1_sol}
f_1 = \frac{\Omega}{6 c R^2} C_1 C_2
	\left[\left(3-\frac{2r}{M}\right)
	\ln N^2 + \frac{2M^2}{3r^2}+\frac{2M}{r}-4\right]
	\mu \ ,
\end{equation}
where $C_2$ is an arbitrary constant to be determined
through the imposition of boundary conditions. Making now
use of the equation (\ref{f1}) we also obtain that
\begin{equation}
\label{f2_sol}
f_2 = -\frac{\Omega}{c R^2} C_1 C_2 N
	\left[\left(1-\frac{r}{M}
	\right)\ln N^2-2-\frac{2M^2}{3r^2N^2}\right] 
	\mu \ .
\end{equation}

	In a similar way, the solutions to equations
(\ref{f3}) and (\ref{f4}) are found to be 
\begin{eqnarray}
\label{f3_sol}
&& f_3 = \frac{15\omega r^3}{16M^5c}\left\{C_3\left[
	\left(3-\frac{2r}{M}\right)
	\ln N^2 + \frac{2M^2}{3r^2}+\frac{2M}{r}-4\right]
	+\frac{2M^2}{5r^2}\ln N^2
	+\frac{4M^3}{5r^3} \right\}\mu \ ,
\\\nonumber\\
\label{f4_sol}
&& f_4 = -\frac{45\omega r^3}{8M^5 c}N\left\{C_3
	\left[\left(1-\frac{r}{M}
	\right)\ln N^2-2-\frac{2M^2}{3r^2N^2}\right] 
	+\frac{4M^4}{15r^4 N^2}\right\}\mu \ ,
\end{eqnarray}

\noindent where again $C_3$ is an arbitrary constant
determined through the boundary conditions. Finally, the
functions $g_3$ and $g_4$ are given by
\begin{eqnarray}
\label{g2_sol}
&& g_2 = \frac{3\Omega r}{8 M^3cN}\left[\ln N^2+
	\frac{2M}{r}\left(1+\frac{M}{r}\right)\right]
	\mu \ ,
\\\nonumber \\
\label{g4_sol}
&& g_4 = -\frac{\omega}{\Omega}g_2 
	= -\frac{3\omega r}{8 M^3cN}\left[\ln N^2+
	\frac{2M}{r}\left(1+\frac{M}{r}\right)\right]
	\mu \ ,
\end{eqnarray}
\noindent so that $g_2+g_4 = ({\bar \omega}/\Omega) g_2$.

	Collecting again all the expressions for the
radial eigenfunctions, the stationary vacuum electric
field external to a misaligned magnetized relativistic
star is given by
\begin{eqnarray}
\label{eeff_1}
&& E^{\hat r} = \Bigg\{
	\frac{15\omega r^3}{16M^5c}\left\{C_3\left[
	\left(3-\frac{2r}{M}\right)
	\ln N^2 + \frac{2M^2}{3r^2}+\frac{2M}{r}-4\right]
	+\frac{2M^2}{5r^2}\ln N^2
	+\frac{4M^3}{5r^3} \right\}
\nonumber\\
	&& \ \ \ \ \ \ \ \ \ \ 
	+ \frac{\Omega}{6 c R^2} C_1 C_2
	\left[\left(3-\frac{2r}{M}\right)
	\ln N^2 + \frac{2M^2}{3r^2}+\frac{2M}{r}-4\right]
	\Bigg\}
	\left[\cos\chi (3 \cos^2\theta-1)+3\sin\chi\cos\lambda
	\sin\theta\cos\theta\right]
	\mu \ , 
\\\nonumber \\
\label{eeff_2}
&& E^{\hat \theta} = - \Bigg\{
	\frac{45\omega r^3}{16 M^5 c}N\left\{C_3
	\left[\left(1-\frac{r}{M}
	\right)\ln N^2-2-\frac{2M^2}{3r^2N^2}\right] 
	+\frac{4M^4}{15r^4 N^2}\right\}
\nonumber\\
	&& \ \ \ \ \ \ \ \ \ \ \ \ 
	+ \frac{\Omega}{2 c R^2} C_1 C_2 N
	\left[\left(1-\frac{r}{M}
	\right)\ln N^2-2-\frac{2M^2}{3r^2N^2}\right] 
	\Bigg\}
	\left[2\cos\chi\sin\theta\cos\theta -
	\left(\cos^2\theta-\sin^2\theta\right)
	\sin\chi\cos\lambda\right] \mu
\nonumber\\
	&& \ \ \ \ \ \ \ \ 
	+\frac{3\bar{\omega} r}{8 M^3 c N}\left[\ln N^2+
	\frac{2M}{r}\left(1+\frac{M}{r}\right)\right] 
	(\sin\chi\cos\lambda) 
	\mu \ ,
\\\nonumber \\
\label{eeff_3}
&& E^{\hat \phi} = - \Bigg\{
	\frac{45\omega r^3}{16 M^5 c}N\left\{C_3
	\left[\left(1-\frac{r}{M}
	\right)\ln N^2-2-\frac{2M^2}{3r^2N^2}\right] 
	+ \frac{4M^4}{15r^4 N^2}\right\}
	+ \frac{\Omega}{2 c R^2} C_1 C_2 N
	\left[\left(1-\frac{r}{M}
	\right)\ln N^2-2-\frac{2M^2}{3r^2N^2}\right] 
\nonumber\\
	&& \ \ \ \ \ \ \ \ \ \ \ \ 
	- \frac{3\bar{\omega} r}{8 M^3 c N}\left[\ln N^2+
	\frac{2M}{r}\left(1+\frac{M}{r}\right)\right]
	\Bigg\}	(\sin\chi\cos\theta\sin\lambda)
	\mu \ .
\end{eqnarray}

	As anticipated in Section \ref{mst_es},
expressions (\ref{eeff_1})--(\ref{eeff_3}) confirm that
the general relativistic dragging of reference frames
introduces a new contribution to the form of the electric
field which does not have a flat spacetime analogue. This
effect is ${\mathcal O}(\omega)$ and therefore present
already in a slow rotation approximation. This is in
contrast with what happens for the magnetic fields, where
higher order approximations of the form of the metric are
necessary for frame dragging corrections to appear.

	The values of the arbitrary constants $C_2$ and
$C_3$ can now be found after imposing the continuity of the
tangential electric field across the star surface. Using
then (\ref{ief_srst_1})--(\ref{ief_srst_3}) as solutions
for the internal electric field and imposing that
$[E^{\hat \theta}(r=R)]_{_{IN}}=[E^{\hat
\theta}(r=R)]_{_{EXT}}$ as well as $[E^{\hat
\phi}(r=R)]_{_{IN}}=[E^{\hat \phi}(r=R)]_{_{EXT}}$, yields
\begin{eqnarray}
\label{c2}
&& C_2 = \frac{1}{N^2_{_R}}\left[\left(1-\frac{R}{M}\right)
	\ln N^2_{_R} - 2 - \frac{2 M^2}{3R^2 N^2_{_R}}\right]^{-1} \ ,
\\ \nonumber \\
&& C_3 = \frac{2M^2}{15 R^2}C_2
	\left[\ln N^2_{_R} + \frac{2M}{R}\right]\ ,
\end{eqnarray}
with $N^2_{_R} \equiv N^2(r=R) = 1 - 2M/R$. It is now
also possible to calculate the surface charge
distribution $\sigma_s$ resulting from the discontinuity
across the star's surface of the radial electric
field. Explicit expressions for this, as well as for the
surface currents corresponding to the discontinuities
across the surface of $B^{\hat \theta}$ and $B^{\hat
\phi}$, will not be given here but can be found in
Appendix C.

	Before concluding this Section on stationary
solutions we will comment on the relevant limits of
equations (\ref{eeff_1})--(\ref{eeff_3}). Firstly, we
verify that they reduce to the Deutsch solutions
(\ref{ed_1})--(\ref{ed_3}) in the limit $\omega=0$ and
$M/r,\ M/R \rightarrow 0$. In this case, in fact
\begin{eqnarray}
&&\lim_{M/r,\ M/R \rightarrow 0} f_1(r) = 
	-\frac{\mu \Omega R^2}{c r^4} =
	\lim_{M/r,\ M/R \rightarrow 0} g_1(r) =
	\lim_{M/r,\ M/R \rightarrow 0} g_5(r)	
	\ , 
\\\nonumber\\
&&\lim_{M/r,\ M/R \rightarrow 0} f_2(r) = 
	-\frac{2 \mu \Omega R^2}{c r^4} \ , 
\\\nonumber\\
&&\lim_{M/r,\ M/R \rightarrow 0} g_2(r) = 
	-\frac{\mu \Omega}{c r^2} \ , 
\\\nonumber\\
&&\lim_{M/r,\ M/R \rightarrow 0} g_4(r) = 
	\frac{\mu \omega}{c r^2} \ , 
\\\nonumber\\
&&\lim_{M/r,\ M/R \rightarrow 0} f_3(r) = 0 =
	\lim_{M/r,\ M/R \rightarrow 0} f_4(r) = 
	\lim_{M/r,\ M/R \rightarrow 0} g_3(r) = 
	\lim_{M/r,\ M/R \rightarrow 0} g_6(r) \ .
\end{eqnarray}

\noindent Secondly, in the limit of an aligned dipole in
a Schwarzschild spacetime, $\chi=0=\omega$, and equations
(\ref{eeff_1})--(\ref{eeff_3}) reduce to
\begin{eqnarray}
\label{eef_sts_1}
&& E^{\hat r} = -\frac{\Omega R}{4 M^3 c N^2_{_R}} 
	\left[\ln N^2_{_R}
	+\frac{2M}{R}\left(1+\frac{M}{R}\right)\right]
	\left[\left(1-\frac{R}{M}\right)
	\ln N^2_{_R} - 2 - \frac{2 M^2}{3R^2 N^2_{_R}}\right]^{-1}
\nonumber \\
	&& \hskip 6.25 cm
	\left[\left(3-\frac{2r}{M}\right)
	\ln N^2 + \frac{2M^2}{3r^2}+\frac{2M}{r}-4\right]
	(3 \cos^2\theta-1)\mu\ , 
\\
\label{eef_sts_2}
&& E^{\hat \theta} = \frac{3\Omega R}{4 M^3 c N^2_{_R}}
	\left[\ln N^2_{_R}
	+\frac{2M}{R}\left(1+\frac{M}{R}\right)\right]
	\left[\left(1-\frac{R}{M}\right)
	\ln N^2_{_R} - 2 - \frac{2 M^2}{3R^2 N^2_{_R}}\right]^{-1}
\nonumber \\
	&& \hskip 5.6 cm
	N \left[\left(1-\frac{r}{M}
	\right)\ln N^2-2-\frac{2M^2}{3r^2N^2}\right] 
	(\sin\theta\cos\theta)\mu \ ,
\\\nonumber \\
\label{eef_sts_3}
&& E^{\hat \phi} = 0 \ .
\end{eqnarray}

\noindent Note that (\ref{eef_sts_1})--(\ref{eef_sts_3})
do not coincide with the corresponding expressions found
by Sengupta (1995). A straightforward calculation would
show that his suggested expressions, while reducing to
the Deutsch solution in the flat spacetime limit, do not
satisfy Maxwell equations. A possible explanation for the
disagreement could be found in the method followed by
Sengupta in his derivation which is not based on the
explicit solution of Maxwell equations. Because of this,
subsequent results obtained on the basis of Sengupta's
expressions for the external electric field (e.g. De
Paolis et al., 1999) should be revisited in terms of
expressions (\ref{eef_sts_1})--(\ref{eef_sts_3}).

\section{Non-Stationary Solutions to Maxwell Equations}
\label{nss}

	In this Section we will drop the assumption of
infinite conductivity which prevented the variation of
the star's magnetic moment and led to the stationary
electromagnetic fields presented in the previous
Sections. Here, on the contrary, we are interested in
time evolving electromagnetic fields and, in particular,
in establishing the general relativistic corrections to
the induction equation. A direct consequence of a finite
conductivity is, in fact, the generation of a time
varying charge density and conduction currents which will
be then responsible for the Ohmic decay. Using Maxwell
equations (\ref{max2a}) and Ohm's laws
(\ref{current1})--(\ref{current4}), we find that the
space charge density $\rho_e = \rho_e(t, r, \theta,
\phi)$ inside the star has a zeroth-order contribution
given by
\begin{eqnarray}
\label{rho}
&& \rho_e = \frac{c e^{-\Lambda}}{16\pi^2\sigma r^2\sin\theta}
	\Bigg\{\left[r\left(\sin\theta B^{\hat \phi} 
	\right)_{,\theta}- rB^{\hat\theta}_{\ ,\phi}\right]_{,r}
\nonumber\\
	&& \hskip 4.0truecm +
	\left[e^{\Phi+\Lambda}B^{\hat r}_{\ ,\phi} - 
	\sin\theta \left(e^{\Phi}r B^{\hat \phi} \right)_{,r}
	\right]_{,\theta} e^{-\Phi}+
	\left[\left(e^{\Phi}r B^{\hat \theta} \right)_{,r}
	- e^{\Phi+\Lambda}B^{\hat r}_{\ ,\theta}\right]_{\ ,\phi}
	e^{-\Phi} \Bigg\} + {\mathcal O}\left(\Omega\right) \ ;
\\
\label{138}
	&& \ \ \ \ \ = 
	\frac{c e^{-\Lambda}}{16\pi^2\sigma}
	\left(\frac{\cos\theta \sin\chi \sin\lambda}{\sin\theta}\right)
	\frac{(G-H)}{r}\Phi_{,r}
	 + {\mathcal O}\left(\Omega\right) \ .
\end{eqnarray}

\noindent where the second expression is the one obtained after using the
ansatz (\ref{ansatz_1})--(\ref{ansatz_3}). It follows from equations
(\ref{evolG}) and (\ref{evolH}) that the zero-order term in equation
(\ref{138}) vanishes, so that the leading contribution is at first order
in $\Omega$.  Using now equations (\ref{max1b}), (\ref{max3c}),
(\ref{max3d}), (\ref{rho}) and Ohm's laws (\ref{current3}),
(\ref{current4}), we obtain the evolution equation for the radial
component of magnetic field

\vbox{
\begin{eqnarray}
\label{ind1}
\frac{\partial B^{\hat r}}{\partial t}&=&
	\frac{c^2 e^{-\Lambda}}{4\pi\sigma r^2\sin\theta}
	\Bigg\{\frac{1}{\sin\theta}
	\left[e^{\Phi+\Lambda}B^{\hat r}_{\ ,\phi} - 
	\sin\theta \left(e^{\Phi}r B^{\hat \phi} \right)_{,r}
	\right]_{,\phi}-\frac{\omega e^{-\Phi}}{4\pi\sigma
	\sin\theta}
	\left[e^{\Phi+\Lambda}B^{\hat r}_{\ ,\phi} - 
	\sin\theta \left(e^{\Phi}r B^{\hat \phi} \right)_{,r}
	\right]_{,\phi\phi}
\nonumber\\
	&&-\left\{\sin\theta 
	\left[\left(e^{\Phi}r B^{\hat \theta} \right)_{,r}
	- e^{\Phi+\Lambda}B^{\hat r}_{\ ,\theta}\right]\right\}
	_{,\theta}- \left\{\sin\theta\left[ 
	\frac{{\omega} r}{4\pi\sigma }
	\left[\left(\sin\theta B^{\hat \phi} \right)_{,\theta}
	- B^{\hat\theta}_{\ ,\phi}\right]\right]_{,r}\right\}_{,\theta}
\nonumber\\
	&&-\frac{{\omega}e^{-\Phi}}{4\pi\sigma}\left\{\sin\theta
	\left[e^{\Phi+\Lambda}B^{\hat r}_{\ ,\phi} - 
	\sin\theta \left(e^{\Phi}r B^{\hat \phi} \right)_{,r}
	\right]_{,\theta}\right\}_{, \theta}
	+\frac{{\Omega}e^{-\Phi}}{4\pi\sigma}
	\Bigg\{e^{\Phi}\left\{\sin\theta\left[r\left(\sin\theta B^{\hat \phi} 
	\right)_{,\theta}- rB^{\hat\theta}_{\ ,\phi}\right]_{,r}
	\right\}_{,\theta}
\nonumber\\
	&&+\left\{\sin\theta\left[e^{\Phi+\Lambda}B^{\hat r}_{\ ,\phi} - 
	\sin\theta \left(e^{\Phi}r B^{\hat \phi} \right)_{,r}
	\right]_{,\theta}\right\}_{, \theta}
	+\left\{\sin\theta
	\left[\left(e^{\Phi}r B^{\hat \theta} \right)_{,r}
	- e^{\Phi+\Lambda}B^{\hat r}_{\ ,\theta}\right]_{\ ,\phi}
	\right\}_{,\theta}\Bigg\} \Bigg\} 
	-\Omega B^{\hat r}_{\ ,\phi} \ .
\end{eqnarray}}

	Similarly, using equations (\ref{max1c}),
(\ref{max3b}), (\ref{max3d}), (\ref{rho}) and Ohm's laws
(\ref{current2}), (\ref{current4}), we obtain the
evolution equation for the polar component of magnetic
field
\begin{eqnarray}
\label{ind2}
\frac{\partial B^{\hat \theta}}{\partial t}&=&
	\frac{c^2 e^{-\Lambda}}
	{4\pi\sigma r}\Bigg\{\left\{e^{-\Lambda}\left[\left(e^{\Phi} r
	B^{\hat\theta}
	\right)_{,r}-{e^{\Phi+\Lambda}}B^{\hat r}_{\ ,\theta}\right]
	\right\}_{,r}
	+\left\{e^{-\Lambda}\left[\frac{{\omega} r}{4\pi\sigma}
	\left[\left(\sin\theta B^{\hat \phi} \right)_{,\theta}
	- B^{\hat\theta}_{\ ,\phi}\right]\right]_{,r}\right\}_{,r}
\nonumber \\
	&& 
	+\frac{1}{4\pi\sigma }\left\{{\omega}e^{-(\Phi+\Lambda)}
	\left[e^{\Phi+\Lambda}B^{\hat r}_{\ ,\phi} - 
	\sin\theta \left(e^{\Phi}r B^{\hat \phi} \right)_{,r}
	\right]_{,\theta}\right\}_{,r}
	-\frac{\Omega}{4\pi\sigma}
	\Bigg\{\left\{e^{-\Lambda}\left[r
	\left(\sin\theta B^{\hat \phi} 
	\right)_{,\theta}- rB^{\hat\theta}_{\ ,\phi}\right]_{,r}
	\right\}_{, r}
\nonumber\\
	&&+\left\{e^{-(\Phi+\Lambda)}\left[e^{\Phi+\Lambda}
	B^{\hat r}_{\ ,\phi} - 
	\sin\theta \left(e^{\Phi}r B^{\hat \phi} \right)_{,r}
	\right]_{,\theta}\right\}_{, r}
	+\left\{e^{-(\Phi+\Lambda)}
	\left[\left(e^{\Phi}r B^{\hat \theta} \right)_{,r}
	- e^{\Phi+\Lambda}B^{\hat r}_{\ ,\theta}\right]_{\ ,\phi}
	\right\}_{, r}\Bigg\} \Bigg\}
\nonumber \\
    	&& 
	-\frac{c^2 e^{\Phi}}{4\pi\sigma r^2\sin^2\theta}\left\{
	\left[\left(\sin\theta B^{\hat\phi} \right)_{,\theta}
	- B^{\hat\theta}_{\; ,\phi}\right]_{,\phi}
	-\frac{{\omega} e^{-\Phi}}{4\pi\sigma}
	\left[\left(\sin\theta B^{\hat \phi} \right)_{,\theta}
	- B^{\hat\theta}_{\ ,\phi}\right]_{,\phi\phi}\right\} 
	-\Omega B^{\hat\theta}_{\ , \phi} \ .
\end{eqnarray}

	Finally, equations (\ref{max1d}), (\ref{max3b}),
(\ref{max3c}) and the Ohm's law (\ref{current2}),
(\ref{current3}) yield the evolution equation for the
toroidal part of the magnetic field
\begin{eqnarray}
\label{ind3}
\frac{\partial B^{\hat\phi}}{\partial t}&=&
	-\frac{c^2 e^{-\Lambda}}{4\pi\sigma r\sin\theta}\left\{ 
	\left\{e^{-\Lambda}\left[e^{\Phi+\Lambda} B^{\hat r}_{\ ,\phi}
	-\sin\theta
	\left(e^{\Phi}r B^{\hat\phi}\right)_{,r}\right]\right\}_{,r}
	- \left\{\frac{{\omega} e^{-(\Phi+\Lambda)}}{4\pi\sigma}
	\left[e^{\Phi+\Lambda} B^{\hat r}_{\ ,\phi}-
	\sin\theta \left(re^{\Phi}B^{\hat \phi}\right)_{,r}
	\right]_{,\phi}\right\}_{, r}\right\}
\nonumber\\
	&& +\frac{c^2e^\Phi}{4\pi\sigma r^2}
	\Bigg\{\left\{\frac{1}{\sin\theta}
	\left[\left(\sin\theta B^{\hat \phi} \right)_{,\theta}
	- B^{\hat\theta}_{\ ,\phi}\right]\right\}_{,\theta}
	-\frac{{\omega}e^{-\Phi}}{4\pi\sigma }\left\{
	\frac{1}{\sin\theta}\left[\left(\sin\theta B^{\hat \phi} 
	\right)_{,\theta}- B^{\hat\theta}_{\ ,\phi}\right]_{,\phi}
	\right\}_{,\theta }\Bigg\} 
\nonumber\\
	&& +\frac{\Omega e^{-\Lambda}\sin\theta}{r}
	\left(r^2B^{\hat r}\right)
	_{\ ,r}+ \Omega\left(\sin\theta B^{\hat\theta}\right)_
	{\ ,\theta} \ .
\end{eqnarray}

	A first important feature of equations
(\ref{ind1})--(\ref{ind3}) is that besides the
relativistic corrections due the monopolar part of the
gravitational field (proportional to $M/R$ and already
present in the non-rotating case), the rotation of
spacetime introduces additional corrections related to
the dipolar part of the gravitational field (and
proportional to $\omega$) to the decay of the magnetic
field. A second relevant aspect of equations
(\ref{ind1})--(\ref{ind3}) is that they do not show the
degeneracy encountered in the time evolution of the
magnetic field in a non-rotating spacetime [cf. equation
(\ref{evolF1})]. In that case, in fact, the three
induction equations for the components of the magnetic
field reduce to a single evolution equation (Sengupta,
1997; Geppert et al. 2000). Here, instead, the three
equations remain distinct and a particular field
component might be favoured during the decay. Finally,
equations (\ref{ind1})--(\ref{ind3}) do not factor out
the angular part as it is the case for a non-rotating,
aligned dipole [cf. equation (\ref{evolF1})] and the
evolution of the magnetic field has therefore properties
which depend on the angular position in the star. As a
consequence of this, an initially dipolar magnetic field
might not remain as such during its decay. This could be
relevant for the evolution of the magnetic field of
pulsars and more particularly of magnetars.

	Using now the ansatz
(\ref{ansatz_1})--(\ref{ansatz_3}), we can write
equations (\ref{ind1})--(\ref{ind3}) in the more compact
form
\begin{eqnarray}
\label{evolF}
&&\frac{\partial F}{\partial t}\left(\cos\theta\cos\chi + 
	\sin\theta\sin\chi\cos\lambda\right)\sin\theta 
	= \frac{c^2 e^{-\Lambda}}{4\pi\sigma r^2}\Bigg\{
	\left[e^{\Phi}r\left(G-H\right)\right]_{,r}\sin\chi\cos\lambda
\nonumber\\	
	& & \hskip 1.0 cm 
	-2\left[\left(e^{\Phi}r G\right)_{,r}
	+e^{\Phi+\Lambda}F\right]\sin\theta\left(
	\cos\theta\cos\chi + \sin\theta\sin\chi\cos\lambda\right)
\nonumber\\	
	& & \hskip 1.0 cm 
	-\frac{1}{4\pi\sigma}\sin\chi\sin\lambda\biggl\{
	\left[\omega r (H-G)\right]_{,r}
	(1 - 2\sin^2\theta) + 
	2{\omega}e^{-\Phi}\left[\left(e^{\Phi}r H\right)_{,r}+
	e^{\Phi+\Lambda}F\right]
	\sin^2\theta 
\nonumber\\	
	& & \hskip 1.0 cm 
	- \Omega r\left(G-H\right)\Phi_{,r}
	\left(1-2\sin^2\theta\right)\biggr\}
	\Bigg\} \ ,
\\ \nonumber\\
\label{evolG}
&&\frac{\partial G}{\partial t}\left(\sin\theta\cos\chi -
        \cos\theta\sin\chi\cos\lambda\right)
        = \frac{c^2}{4\pi\sigma r}\Bigg\{
        \frac{e^{\Phi}\left(G-H\right)}{r\sin^2\theta}
        \cos\theta\sin\chi\left[\cos\lambda+
        \frac{{\omega}e^{-\Phi}}{4\pi\sigma}\sin\lambda\right]
\nonumber\\
        & & \hskip 1.0 cm
        +e^{-\Lambda}\left[e^{-\Lambda}\left(e^{\Phi}r G\right)_{,r}
        +{e^{\Phi}}F\right]_{,r}\left(
        \sin\theta\cos\chi - \cos\theta\sin\chi\cos\lambda\right)
\nonumber\\
        & & \hskip 1.0 cm
        -\frac{e^{-\Lambda}}{4\pi\sigma}\cos\theta\sin\chi\sin\lambda
        \bigg\{
        e^{-\Lambda}\left[\omega r
        \left(G-H\right)\right]_{,r}
        +\omega\left[F+e^{-(\Lambda+\Phi)}
        \left(r e^{\Phi} H\right)_{,r}\right]
        + \Omega\left[\Phi_{,r}e^{-\Lambda}r\left(G-H\right)\right]
        \bigg\}_{,r}\Bigg\} \ ,
\nonumber\\ \\ \nonumber\\
\label{evolH}
&&\frac{\partial H}{\partial t}\sin\lambda =
	 \frac{c^2 e^{-\Lambda}}{4\pi\sigma r}
	\left\{\left[e^{-\Lambda}\left(e^{\Phi}r H\right)_{,r}
	 +e^{\Phi}F\right]\left[\sin\lambda -
	\frac{{\omega}e^{-\Phi}}{4\pi\sigma}
	\cos\lambda \right]\right\}_{, r} 
	 +\frac{c^2e^\Phi \left(G-H\right)}{4\pi\sigma r^2
	\sin^2\theta}\left[\sin\lambda -
	\frac{{\omega}e^{-\Phi}}{4\pi\sigma}
	\cos\lambda \right]\ ,
\end{eqnarray}
where $F$, $G$, and $H$ satisfy the constraint equation
(\ref{max1a})
\begin{equation}
\left[\left(r^2 F\right)_{, r} + 2 e^{\Lambda}r G\right] 
	\sin\theta\left(\cos\chi\cos\theta +
	\sin\chi\sin\theta\cos\lambda\right)+
	e^{\Lambda}r \left(H-G\right)\sin\chi\cos\lambda = 0 \ ,
\end{equation}
The set of equations (\ref{evolF})--(\ref{evolH}) is too
complicated to be solved analytically even when analytic
expressions are available for the metric functions
(e.g. for a constant density stellar model). The
numerical solution of (\ref{evolF})--(\ref{evolH}) for a
number of equations of state together with a
self-consistent evolution of the star's angular velocity
and electrical conductivity will be presented in a
separate paper (Rezzolla et al. 2000). Note that
equations (\ref{evolF})--(\ref{evolH}) could also be
derived through a vector potential $A_{\mu}$ defined so
that the electromagnetic tensor $F_{\mu \nu}= A_{\nu ,
\mu} - A_{\mu , \nu}$. Details of this derivation can be
found in Appendix D.
	
	An interesting limit of the induction equations
(\ref{evolF}), (\ref{evolG}) is the one for a
non-rotating dipole in a spherically symmetric
spacetime. In this case: $\Omega = 0 = \omega$, $\chi=0$,
and $H, H /\partial t$ are not determined [cf. equations
(\ref{ansatz_3}) and (\ref{psi3})]. As mentioned before,
the induction equations are degenerate in this case and
the unique evolution equation is then
\begin{eqnarray}
\label{evolF1}
&&\frac{\partial F}{\partial t}	= \frac{c^2 e^{-\Lambda}}
	{4\pi\sigma r^2}\left\{\left[
	e^{\Phi-\Lambda}\left(r^2 F\right)_{,r}\right]_{,r}
	- 2 e^{\Phi+\Lambda}F\right\} \ ,
\end{eqnarray}
corresponding to the solution found by Geppert et al.
(2000). When the metric functions $\Phi$ and $\Lambda$
refer only to the vacuum region of spacetime external to
the star, equation (\ref{evolF1}) further simplifies to
\begin{eqnarray}
\label{evolF2}
&&\frac{\partial F}{\partial t}	= \frac{c^2}{4\pi\sigma r^2}
	\sqrt{\frac{r-2M}{r}}\left\{\left[
	\left(1 - \frac{2M}{r}\right)\left(r^2 F\right)_{,r}\right]_{,r}
	- 2 F\right\} \ ,
\end{eqnarray}
and which now corresponds to the solution found by
Sengupta (1997).

\section{Conclusion}
\label{conclusion}
	We have presented analytic general relativistic
expressions for the electromagnetic fields internal and
external to a slowly-rotating magnetized neutron star.
The star is considered isolated and in vacuum, but no
special assumption is made on the orientation of the
dipolar magnetic field with respect to the rotation axis.
The solutions to Maxwell equations have been considered
both for an infinite and for a finite electrical
conductivity.

	In the first case, corresponding to stationary
magnetic fields, we have shown that the general
relativistic corrections due to the dragging of reference
frames are not present in the form of the magnetic fields
but emerge only in the form of the electric fields. In
particular, we have shown that the frame-dragging
provides an additional induced electric field which is
analogous to the one introduced by the rotation of the
star in the flat spacetime limit. In the case of finite
electrical conductivity, on the other hand, corresponding
to decaying magnetic fields, we have shown that
corrections due both to the spacetime curvature and to
the dragging of reference frames can be found in the
induction equation. An interesting result obtained in
this regime is that the rotation of the star eliminates
the degeneracy in the components of the induction
equation which remain therefore distinct. Furthermore,
rotation and dipole misalignment do not eliminate the
angular dependence in the induction equation and, as a
result, an initially dipolar magnetic field might evolve
towards a different configuration during its decay.

	Because of their complexity, the evolution
equations found for the magnetic field require a
numerical integration which will discuss in detail in a
forthcoming work (Rezzolla et al. 2000). There, we will
also present direct comparisons between the flat and the
curved spacetime solutions and quantify more precisely
the importance of the general relativistic corrections.

	One of the relevant aspects of the solutions
presented in this paper is that they provide a lowest
order analytic form for the electromagnetic field in the
spacetime of a slowly rotating misaligned dipole subject
to assumptions which, while giving simplifications, allow
the major features of a realistic solution to be seen. In
this sense, they reflect a rather general physical
configuration and could therefore be used in a variety of
astrophysical situations.

\section*{Acknowledgments}

We gratefully acknowledge the referee, Ulrich Geppert,
for carefully reading the manuscript. Financial support
for this work is provided by the Italian Ministero
dell'Universit\`a e della Ricerca Scientifica e
Tecnologica. B.A. is grateful to The Abdus Salam
International Centre for Theoretical Physics, where part
of this research was carried out.

\appendix

\section[]{The electromagnetic tensor}

	For completeness, we provide below the explicit
expressions for the components of the electromagnetic
tensor used throughout the paper. In a coordinate basis,
and at first order in $\Omega$, these components are
given by
\begin{equation}
\label{fab}
F_{\alpha \beta} = \left( 
\begin{array}{cccc}
0 
	&-e^{\Phi} E_{r}  
	-\omega
	e^{\Lambda} r^2\sin\theta B^{\theta}
	&-e^{\Phi}E_{\theta}
	+\omega
	e^{\Lambda}r^2\sin\theta B^{r}
	&-e^{\Phi} E_{\phi} \\ \\ 
e^{\Phi} E_{r} + \omega 
	e^{\Lambda} r^2\sin\theta B^{\theta}
	& 0 
	& e^{\Lambda}r^2\sin\theta B^{\phi} 
	& -e^{\Lambda}r^2\sin\theta B^{\theta} \\ \\ 
e^{\Phi}E_{\theta}
	-\omega e^{\Lambda}r^2\sin\theta B^{r}
	& -e^{\Lambda}r^2\sin\theta B^{\phi} 
	& 0 
	& e^{\Lambda}r^2\sin\theta B^{r} \\ \\ 
e^{\Phi} E_{\phi}
	& e^{\Lambda}r^2\sin\theta B^{\theta} 
	& -e^{\Lambda}r^2\sin\theta B^{r}  
	& 0 \\ \\ 
\end{array}
\right) \ .
\end{equation}

	The matrix (\ref{fab}) can also be expressed in
terms of the electromagnetic field measured by the ZAMO
observers, in which case it takes the form

\begin{equation}
F_{\alpha \beta} = \left( 
\begin{array}{cccc}
0 
	&-e^{\Phi+\Lambda} E^{\hat r} - 
	\omega
	e^{\Lambda} r\sin\theta B^{\hat \theta}
	&-\displaystyle e^{\Phi} r E^{\hat \theta}
	+\omega r^2\sin\theta B^{\hat r}
	&-e^{\Phi} r \sin\theta E^{\hat \phi} \\ \\ 
e^{\Phi+\Lambda} E^{\hat r} +\omega
	e^{\Lambda} r\sin\theta B^{\hat \theta} 
	& 0 
	& e^{\Lambda}r B^{\hat\phi} 
	& -e^{\Lambda}r\sin\theta B^{\hat\theta} \\ \\ 
\displaystyle e^{\Phi} r E^{\hat \theta}
	-\omega r^2\sin\theta B^{\hat r} 
	& -e^{\Lambda}r B^{\hat\phi} 
	& 0 
	& r^2\sin\theta B^{\hat r} \\ \\ 
e^{\Phi} r \sin\theta E^{\hat \phi} 
	& e^{\Lambda}r\sin\theta B^{\hat\theta} 
	& -r^2\sin\theta B^{\hat r}  
	& 0 \\ \\ 
\end{array}
\right) \ .
\end{equation}

	Finally, we note that the components of the electromagnetic
tensor in the ZAMO frame can be derived from
(\ref{fab}) with the transformation

\begin{equation}
F_{{\hat \alpha} {\hat \beta}} = 
	{\boldsymbol e}_{\hat \alpha}^{\mu}
	{\boldsymbol e}_{\hat \beta}^{\nu} F_{\mu \nu} \ ,
\end{equation}
to obtain

\begin{equation}
F_{{\hat \alpha} {\hat \beta}} = \left( 
\begin{array}{cccc}
0 
	&-c E^{\hat r} 
	&-c E^{\hat \theta}
	&-c E^{\hat \phi} \\ \\ 
c E^{\hat r}
	& 0 
	& B^{\hat \phi} 
	& -B^{\hat \theta} \\ \\ 
c E^{\hat \theta}
	& -B^{\hat \phi} 
	& 0 
	& B^{\hat r} \\ \\ 
c E^{\hat \phi}
	& B^{\hat \theta} 
	& -B^{\hat r}  
	& 0 \\ \\ 
\end{array}
\right) \ .
\end{equation}

\section[]{Radial Eigenfunctions}

	In this Appendix we briefly sketch the procedure
for the calculation of the radial eigenfunctions and that
have lead to the solutions
(\ref{f_of_r})--(\ref{h_of_r}),
(\ref{f1_sol})--(\ref{f2_sol}),
(\ref{g2_sol})--(\ref{g4_sol}). In general, we look for a
solution of the equation 
\begin{equation}
\label{leg_eq_g}
\frac{d}{dx}\left\{\left(\frac{1+x}{1-x}\right)\frac{d}{dx}
	\left[\left(1-x\right)^2 q_{\ell}\right]\right\}+
	\ell\left(\ell+1\right)	q_{\ell} = 0 \ ,
\end{equation}
for the function $q_{\ell}$. Equation (\ref{leg_eq_g})
can also be written as
\begin{equation}
\label{leg_eq_expl}
\left(1-x^2\right)q''_{\ell}- 2(1+2x)q'_{\ell} + 
	\left[\ell\left(\ell+1\right) - 2 \right] q_{\ell} = 0 \ ,
\end{equation}
with the dash representing a total derivative with
respect to $x$. The solution of (\ref{leg_eq_g}) has then
form
\begin{equation}
\label{q_ell}
q_{\ell}=\frac{d}{dx}\left[\left(1+x\right)\frac{d}{dx}Q_{\ell}\right]
	\ ,
\end{equation}
where $Q_{\ell}$ are Legendre functions of second kind
(Jeffrey 1995). A proof of this comes from substituting
(\ref{q_ell}) into (\ref{leg_eq_expl}) to obtain
\begin{equation}
(1+x)\left[(1-x^2)q''''_{\ell} + (1-7x)q'''_{\ell} +
	\left[\ell\left(\ell+1\right)- 2\right]q''_{\ell}\right]
	- 4(1+2x)q''_{\ell} +  \left[\ell\left(\ell+1\right)-
	2\right]q'_{\ell} = 0 \ ,
\end{equation}
and which can be rewritten as 
\begin{equation}
\label{leg_full}
(1+x)\left[(1-x^2)q''''_{\ell} -6xq'''_{\ell} +
	\left[\ell\left(\ell+1\right)- 6\right]q''_{\ell}\right]
	+(1-x^2)q'''_{\ell} - 4xq''_{\ell} +  \left[\ell\left(\ell+1\right)-
	2\right]q'_{\ell} = 0 \ .
\end{equation}

	It is now easy to realize that (\ref{leg_full})
is identically satisfied since the content of the square
brackets is, in fact, the second derivative of Legendre's
equation
\begin{equation}
\label{leg_eq}
(1-x^2)q''_{\ell}- 2xq'_{\ell} + 
	\ell\left(\ell+1\right) q_{\ell} = 0 \ ,
\end{equation}
while all the remainder of (\ref{leg_full}) is the first
derivative of Legendre's equation (\ref{leg_eq}).

\section[]{Surface Charges and Currents}

	We here give explicit expressions for the surface
charge distribution $\sigma_s$ and surface currents
$i^{\hat k}$ resulting from the discontinuities across
the star's surface of the $r-$component of the electric
field and of $\theta,\ \phi-$components of the magnetic
field. 

	Defining now $[A]^{\pm}\equiv [A(r=R)]_{_{EXT}} -
[A(r=R)]_{_{IN}}$, the surface charge density $\sigma _s$
can be found as
\begin{equation}
\sigma_s = \frac{1}{4\pi} [E^{\hat r}]^{\pm}\ ,
\end{equation}
and is given by
\begin{eqnarray}
\sigma_s &=& \frac{1}{4\pi}\Bigg\{
	\frac{15\omega_R R^3}{8M^5c}\left\{C_3\left[
	\left(3-\frac{2R}{M}\right)
	\ln N_R^2 + \frac{2M^2}{3R^2}+\frac{2M}{R}-4\right]
	+\frac{2M^2}{5R^2}\ln N_R^2
	+\frac{4M^3}{5R^3} \right\}
\nonumber\\
	&& \ \ \ \ \ \ \
	+ \frac{\Omega}{3 c R^2} C_1 C_2
	\left[\left(3-\frac{2R}{M}\right)
	\ln N_R^2 + \frac{2M^2}{3R^2}+\frac{2M}{R}-4\right]
	\Bigg\}(\cos\chi) \mu
\nonumber\\
	&& 
	\!\!\!\!\!
	-\frac{1}{4\pi}\Bigg\{
	\frac{45\omega_R R^3}{16M^5c}\left\{C_3\left[
	\left(3-\frac{2R}{M}\right)
	\ln N_R^2 + \frac{2M^2}{3R^2}+\frac{2M}{R}-4\right]
	+\frac{2M^2}{5R^2}\ln N_R^2
	+\frac{4M^3}{5R^3} \right\}
\nonumber\\
	&& \ \ \ \ \ \ \ \ \
	+ \frac{\Omega}{2 c R^2} C_1 C_2
	\left[\left(3-\frac{2R}{M}\right)
	\ln N_R^2 + \frac{2M^2}{3R^2}+\frac{2M}{R}-4\right]
	-\frac{{\bar\omega}_R}{cR^2}C_1	\Bigg\}
	\sin\theta (\cos\chi\sin\theta-
	\sin\chi\cos\lambda\cos\theta)\mu \ ,
\end{eqnarray}	
where ${\omega}_R\equiv {\omega}(r=R)$ and
${\bar\omega}_R\equiv {\bar \omega}(r=R)$

	In a similar way, imposing that 
\begin{equation}
i^{\hat\phi} = \frac{c}{4\pi}[B^{\hat\theta}]^{\pm}\ , 
	\hskip 3.0 cm
i^{\hat\theta} = \frac{c}{4\pi}[B^{\hat\phi}]^{\pm}\ ,
\end{equation}
we obtain

\begin{eqnarray}
&& i^{\hat \theta} = \frac{3c}{16 \pi}\frac{N_R}{M^2 R}
	\left[\frac{R}{M}\ln N_R^2 +\frac{1}{N_R^2}+ 1+
	\frac{4M^2}{3R^2}C_1\right]
	\left(\sin\chi \sin\lambda\right)\mu
	\ ,
\\ \nonumber \\
&& i^{\hat \phi} = \frac{3c}{16 \pi}\frac{N_R}{M^2 R}
	\left[\frac{R}{M}\ln N_R^2 +\frac{1}{N_R^2}+ 1+
	\frac{4M^2}{3R^2}C_1\right]\left(\cos\chi \sin\theta
	- \sin\chi \cos\theta \cos\lambda \right)
	\mu \ .
\end{eqnarray}

\section[]{An Alternative Derivation of the Induction Equation}

	To confirm the results presented in Section
\ref{nss} and to compare with the results presented in
the literature in the case of Schwarzschild background
spacetime (Sengupta, 1997) we here present a derivation
of equations (\ref{evolF}), (\ref{evolG}), (\ref{evolH})
in terms of a vector potential $A_{\alpha}$ defined as
\begin{equation}
\label{a_def}
F_{\alpha \beta} \equiv 
	A_{\beta,\alpha} - A_{\alpha,\beta} \ .
\end{equation}
The use of a vector potential is sometimes looked at with
skepticism (Geppert, Page and Zannias, 2000) in view of
the non-commutativity of the covariant derivative, which
could lead to ambiguities if Maxwell equations are
expressed through double covariant derivatives of a
vector potential. All of these ambiguities, however, are
easily removed if the vector potential is introduced only
when Maxwell equations (\ref{maxwell_secondpair}) are
recast in a form not involving covariant derivatives
\begin{equation}
\label{maxwell_secondpair_sr}
\frac{1}{\sqrt{-g}}
	\left(\sqrt{-g}F^{\alpha\beta}\right)_{,\beta} 
	= {4\pi} J^{\alpha} 
\end{equation}

	In our derivation we start by using equations
(\ref{maxwell_firstpair}), (\ref{maxwell_secondpair}),
(\ref{ohm}), neglecting the displacement current and
taking the four-velocity of the conductor $u^\alpha$ in
the form (\ref{vel}) [i.e. neglecting terms proportional
to $g^{a0}F_{0 b}u^b\approx {\mathcal O}(\omega^2)$] we
get
\begin{equation}
\frac{1}{4\pi\sigma\sqrt{-g}u^0}\left(\sqrt{-g}
	F^{a b}\right)_{,b} = \rho_e u^a
	+\sigma g^{a b}\left(F_{b c}u^c +F_{b 0}u^{0}\right) \ ,
\end{equation}
which can be written as
\begin{equation}
F_{a 0}=\frac{g_{a b}}{4\pi\sigma\sqrt{-g}u^0}
	\left(\sqrt{-g}F^{b c}\right)_{,c}-
	\frac{F_{a b}u^b}{u^0}-\frac{\rho_e g_{a b}u^b}{\sigma u^0} \ .
\end{equation}
Using now (\ref{maxwell_secondpair_sr}) we can write the
general expression for the evolution of the vector
potential $A_i$ as
\begin{equation}
\label{ai0}
A_{i,0}= - F_{i 0} = -\frac{g_{i j}}{4\pi\sigma\sqrt{-g}u^0}
	\left(\sqrt{-g}F^{j k}\right)_{,k}+
	\frac{F_{i j}u^j}{u^0}+\frac{\rho_e g_{ij}u^j}{\sigma u^0}\ .
\end{equation}
	
	Using (\ref{ai0}), the induction equation for the
evolution of the $\phi$ component of vector potential can
be written as
\begin{eqnarray}
\label{evolA_p}
\frac{\partial A_\phi}{\partial t} &=& -\frac{c^2 e^{-\Lambda}}{4\pi\sigma}
        \sin\theta\Bigg\{\frac{1}{\sin\theta}\left(e^{\Phi-\Lambda}
        F_{\phi r}\right)_{,r} +\left(\frac{e^{\Phi+\Lambda}F_{\phi\theta}}
        {r^2\sin\theta}\right)_{,\theta}
        + \frac{{\omega}e^{-\Phi}}{4\pi\sigma}
        \left[\sin\theta\left(e^{\Phi-\Lambda}F_{\theta r}\right)_{,r}+
        \frac{e^{\Phi+\Lambda}}{r^2\sin\theta }
        F_{\theta\phi ,\phi}\right]_{,\theta}
\nonumber\\
        & & + \frac{1}{4\pi\sigma}\left\{\omega e^{-\Lambda}
        \left[\left(\sin\theta
        F_{r\theta}\right)_{,\theta}+\frac{1}{\sin\theta}
        F_{r\phi ,\phi}\right]\right\}_{,r}\Bigg\}
        +\frac{c^2\Omega e^{-2\Lambda}}{16\pi^2\sigma^2}\Phi_{,r}
  \left[\frac{1}{\sin\theta}F_{\phi r,\phi}+\left(\sin\theta
        F_{\theta r}\right)_{,\theta}\right]\sin\theta \ ,
\end{eqnarray}
which coincides with equation (11) of Sengupta's 1998
paper when $\chi=0=\omega$ and $e^{2\Phi} = N^2 = 1 -
2M/r$. Similarly, the evolution equations for the other
components of the vector potential are given by
\begin{eqnarray}
\label{evolA_t}
\frac{\partial A_\theta}{\partial t} &=& -\frac{c^2 e^{-\Lambda}}{4\pi\sigma 
	\sin\theta}\Bigg\{{\sin\theta}\left(e^{\Phi-\Lambda}
	F_{\theta r}\right)_{,r}
	+\frac{e^{\Phi+\Lambda}}{r^2\sin\theta }
	F_{\theta\phi , \phi}
	- \frac{{\omega}e^{-\Phi}}{4\pi\sigma} 
	\left[\sin\theta\left(e^{\Phi-\Lambda}F_{\theta r}\right)_{,r\phi}+
	\frac{e^{\Phi+\Lambda}}{r^2\sin\theta }
	F_{\theta\phi ,\phi\phi}\right]\Bigg\}+\Omega
	F_{\theta\phi} \ , 
\\ \nonumber\\
\label{evolA_r}
\frac{\partial A_r}{\partial t} &=& -\frac{c^2 e^{\Lambda}}{4\pi\sigma 
	r^2\sin\theta}
	\Bigg\{e^{\Phi-\Lambda}
	\left(\sin\theta F_{r\theta}\right)_{,\theta} +
	\frac{e^{\Phi-\Lambda}}{\sin\theta}F_{r\phi , \phi}
	- \frac{{\omega}e^{-\Lambda}}{4\pi\sigma} 
	\left[\left(\sin\theta 	F_{r\theta}\right)_{,\theta\phi}
	 +\frac{1}{\sin\theta}F_{r\phi ,\phi\phi}
	\right]\Bigg\}+\Omega F_{r\phi} \ . 
\end{eqnarray}
	In the case of a misaligned rotator, the explicit
expressions for the ``magnetic'' components of the
electromagnetic tensor are
\begin{eqnarray}
&& F_{r\theta} = e^{\Lambda}r H\sin\chi\sin\lambda\ ,
\\ \nonumber\\
&& F_{\phi r} = G e^{\Lambda}r \sin\theta\left(\sin\theta\cos\chi
	-\sin\chi\cos\lambda\cos\theta\right)\ ,    
\\ \nonumber\\
&& F_{\theta\phi} = Fr^2\sin\theta\left(\cos\theta\cos\chi
	+\sin\chi\cos\lambda\sin\theta\right)\ ,
\end{eqnarray}

\noindent using which the induction equations
(\ref{evolA_p})--(\ref{evolA_r}) assume the form
\begin{eqnarray}
\label{vector3}
&& \frac{\partial A_\phi}{\partial t} = \frac{c^2 e^{-\Lambda}\sin\theta}
	{4\pi\sigma}\Bigg\{- \left[\left(e^{\Phi}r G\right)_{,r} 
	+e^{\Phi+\Lambda}F\right]
	\left(\sin\theta\cos\chi - \sin\chi\cos\lambda\cos\theta
	\right)
\nonumber\\
	&&\hskip 1.0truecm 
	+\frac{1}{4\pi\sigma}\left\{
	\left[\left(e^{\Phi}r H\right)_{,r} 
	+e^{\Phi+\Lambda}F\right]{\omega}e^{-\Phi}
	+\left[\omega r(G-H)\right]_{, r}+\Omega r
	\left(G-H\right)\Phi_{,r}\right\}
	\cos\theta\sin\chi\sin\lambda\Bigg\} \ ,
\\ \nonumber\\
\label{vector2}
&& \frac{\partial A_\theta}{\partial t}= 
	\frac{c^2}{4\pi\sigma}
	\left[e^{-\Lambda} \left(e^\Phi r H\right)_{,r}
	+{e^{\Phi}F}\right]
	\left[\sin\chi\sin\lambda -\frac{{\omega}e^{-\Phi}}{4\pi\sigma}
	\sin\chi\cos\lambda\right]
	+\Omega Fr^2\sin\theta\left(\cos\theta\cos\chi
	+\sin\chi\cos\lambda\sin\theta\right) \ , 
\\ \nonumber\\
\label{vector1}
&& \frac{\partial A_r}{\partial t} = 
	\frac{c^2 e^{\Phi+\Lambda}\left(G-H\right)}
	{4\pi\sigma r\sin\theta}
	\sin\chi\cos\theta\left[\sin\lambda -
	\frac{{\omega}e^{-\Phi}}{4\pi\sigma}\cos\lambda\right]
	-\Omega G e^{\Lambda}r \sin\theta \left(\sin\theta\cos\chi
	-\sin\chi\cos\lambda\cos\theta\right) \ . 
\end{eqnarray}
	
	Using now the definition (\ref{a_def}) it is
possible to show that equations (\ref{vector3}),
(\ref{vector2}), (\ref{vector1}) are equivalent to
equations (\ref{evolF}), (\ref{evolG}) and (\ref{evolH})
derived in the main text.

\label{lastpage}

\end{document}